\documentclass[12pt]{article}
\usepackage{amsfonts,amsmath}
\usepackage{amssymb}
\usepackage[T2A]{fontenc}
\usepackage[hypertex] {hyperref}

\newcommand{\Hp}{\Sp^+}

\newcommand{\by}{\bar{y}}

\newcommand{\T}{\mathcal{T}}
\newcommand{\Sp}{{\mathcal H}}
\newcommand{\ust}{\bar{\,*\,}}
\newcommand{\dr}{{{\rm d}}}

\renewcommand{\theequation}{\thesection.\arabic{equation}}
\renewcommand{\thesubsection}{\arabic{section}.\arabic{subsection}}
\makeatletter \@addtoreset{equation}{section} \makeatother

\newcommand{\gt}{\tau}
 \newcommand{\eq}{\eqref}
\newcommand{\ie}{{\it i.e.,} }
{\vspace{3mm} }

\def\al{\alpha}

\def\*{\star}

\def\E2{\mathbf{E}}

\newcommand{\rhs}{{\it r.h.s.} }
\newcommand{\rhss}{{\it r.h.s.'s} }
\newcommand{\lhs}{{\it l.h.s.} }
\newcommand{\be}{\begin{equation}}
\newcommand{\ee}{\end{equation}}
\newcommand{\bee}{\begin{eqnarray}}
\newcommand{\beee}{\begin{array}}
\newcommand{\eee}{\end{eqnarray}}
\newcommand{\eeee}{\end{array}}
\newcommand{\ga}{\alpha}
\newcommand{\gb}{\beta}

\newcommand{\gga}{\gamma}

\newcommand{\gd}{\delta}

\newcommand{\gep}{\epsilon}
\newcommand{\gvep}{\varepsilon}

\newcommand{\go}{\omega}

\newcommand{\dal}{\dot \alpha}

\newcommand{\q}{\,,\qquad}
\newcommand{\nn}{\nonumber}

\newcommand{\p}{\partial}

\newcommand{\ff}{\frac}
\newcommand{\hmt}{\vartriangle}

\begin{document}

\begin{flushright}
FIAN/TD/18-2020\\
\end{flushright}

\vspace{0.5cm}
\begin{center}
{\large\bf Spin-Locality of $\eta^2$ and $\bar\eta^2$ Quartic
Higher-Spin Vertices}

\vspace{1 cm}

\textbf{V.E.~Didenko${}^1$, O.A.~Gelfond${}^{1,2}$, A.V.~Korybut${}^1$ and  M.A.~Vasiliev${}^{1,3}$}\\

\vspace{1 cm}

\textbf{}\textbf{}\\
 \vspace{0.5cm}
 \textit{${}^1$ I.E. Tamm Department of Theoretical Physics,
Lebedev Physical Institute,}\\
 \textit{ Leninsky prospect 53, 119991, Moscow, Russia }\\

\vspace{0.7 cm}\textit{
${}^2$ Federal State Institution "Scientific Research Institute for System Analysis
of the Russian Academy of Science",}\\
\textit{Nakhimovsky prospect 36-1, 117218, Moscow, Russia}

\vspace{0.7 cm} \textit{ ${}^3$
Moscow Institute of Physics and Technology, Institutsky pereulok 9, 141701, Dolgoprudny, Moscow region, Russia}

\vspace{0.7 cm} didenko@lpi.ru, gel@lpi.ru, akoribut@gmail.com, vasiliev@lpi.ru

\par\end{center}

\begin{center}
\vspace{0.6cm}

\par\end{center}

\vspace{0.4cm}

\begin{abstract}
\noindent
Higher-spin theory contains a complex coupling parameter $\eta$.
Different higher-spin vertices are associated with
 different powers of $\eta$ and  its complex conjugate $\bar \eta$.
Using $Z$-dominance Lemma of \cite{2a1}, that controls spin-locality
of the higher-spin equations, we show that the third-order
contribution to the zero-form $B(Z;Y;K)$ admits a $Z$-dominated form
that leads to spin-local
vertices in the  $\eta^2$ and $\bar \eta^2$ sectors of the higher-spin equations.
These vertices include, in particular, the $\eta^2$ and $\bar \eta^2$ parts of the
$\phi^4$ scalar field vertex.
\end{abstract}
\begin{flushright}
{\it In memory of Dima Polyakov}
\end{flushright}

\newpage

\tableofcontents

\bibliographystyle{hieeetr}

\newpage
\section{Introduction}

Higher-spin (HS) gauge theory  is a
theory of an infinite set of gauge fields of all spins. Since
gauge invariant HS interaction vertices contain higher derivatives
of degrees increasing with spin
\cite{Bengtsson:1983pd,Berends:1984wp,Fradkin:1987ks,Fradkin:1991iy},
HS gauge theory is not a local
field theory in the usual sense. Some arguments that HS gauge
theory has to be essentially non-local were given in
\cite{Sleight:2017pcz} based on the holographic correspondence
with the boundary (critical) sigma-model, conjectured by Klebanov
and Polyakov \cite{Klebanov:2002ja} (see also \cite{Sezgin:2002rt}).

To be free from the assumptions of holographic correspondence it
is   important to analyze the issue of (non)locality of HS gauge
theory directly in the bulk. Based on the nonlinear HS equations of
\cite{Vasiliev:1990en,more}  such analysis was performed in
\cite{Vasiliev:2016xui,Gelfond:2017wrh,Vas,4a1, 4a2}
 in different
sectors of the theory at some lowest orders.
All vertices derived in  these papers turned out to
be spin-local\footnote{Roughly speaking, spin-locality implies that the vertices are local for
any finite subset of fields of different spins. More precisely, this is literally the
case at the lowest interaction order but may need some further elaboration at higher orders.
For more detail on these issues we refer the reader to \cite{2a2}.  } including some of the quintic vertices in the
Lagrangian counting. Also somewhat different arguments pointing out at
locality of the HS gauge theory were presented   in a recent paper
\cite{David:2020ptn}.

The obtained vertices agree with holographic prediction at cubic
order \cite{Sezgin:2017jgm, Didenko:2017lsn}. However, the bulk
vertices derived from nonlinear HS equations so far did not
contain the $\phi^4$ vertex for the spin-zero field $\phi$. On the
other hand, it is this vertex \cite{Bekaert:2015tva} that was
argued to be highly non-local \cite{Sleight:2017pcz} in the HS
theory holographically dual to the boundary sigma-model
\cite{Klebanov:2002ja}. The degree of non-locality prescribed by
the analysis of \cite{Sleight:2017pcz} led the authors to a
conclusion of a fundamental failure for HS holographic
reconstruction programme beyond cubic order. Still, the same
vertex was also analyzed in \cite{Ponomarev:2017qab} concluding
that the non-locality if present is of a very special form.

The aim of this paper is to carry out holographically independent
approach to the locality problem. We do it by extending the
analysis of locality of \cite{Vas, 2a1, 4a1, 4a2, 2a2} to the
vertices of order $C^3$ in the sector of equations on the
zero-forms $C$ that contain in particular the $\phi^4$  vertex of
interest in the form of $\phi^3$ contribution to the field
equations. Note that the vertices studied in this paper include an
$AdS_4$ extension of those obtained by Metsaev in
\cite{Metsaev:1991mt}.

In this paper we give general arguments based on the so-called
$Z$-dominance Lemma of \cite{2a1} that the holomorphic, \ie
$\eta^2$, and antiholomorphic $\bar\eta^2$ vertices in HS gauge
theory must be spin-local, where $\eta$ is a complex parameter in
the HS equations. We explicitly demonstrate by direct calculation
that every individual contribution to the (anti)holomorphic part
of a quartic vertex acquires a form that results in a complete
spin-locality of the entire piece. While our analysis is sufficient to
see that the result is local it does not give directly the  manifestly
local form of the remaining local vertex.
The derivation of the latter, which  uses  partial integrations and
Schouten identity, we leave for the future. The analysis of the
mixed $\eta\bar \eta$ vertex that is the only remaining sector in
the analysis of locality of the $\phi^4$ sector, needs somewhat
different tools and is beyond the scope of this paper.

The paper is organized as follows. In Section \ref{Section2}, the
necessary background on HS equations
is presented. Section \ref{Section3} contains brief
recollection on the so called limiting shifted homotopy and the
interpretation of the $Z$-dominance lemma via space $\Hp$
of star-product functions. In
Section \ref{RES}, we collect expressions for the holomorphic
vertices obtained from the generating equations.
Discussion of the obtained results and problems yet to be solved
is placed in Section \ref{Conclussion}. Some useful formulas are collected in Appendix A.
 Appendices    B  and C
   contain the detailed derivation of the third-order
contribution
$B_3^{\eta\eta}$ to the zero-form and second-order contribution
$W_2^{\eta\eta}$ to the one-form fields, respectively.

\section{Recollection of higher-spin equations}\label{Section2}
In the frame-like formalism \cite{Vasiliev:1980as}, unfolded equations for interacting HS fields in $AdS_4$
can be schematically put into the  form \cite{Vasiliev:1988sa}
\begin{equation}\label{HSsketch1}
\dr_x \omega+\go\ast \go=\Upsilon(\go,\go,C)+\Upsilon(\go,\go,C,C)+\ldots,
\end{equation}
\begin{equation}\label{HSsketch2}
\dr_x C+\omega \ast C-C\ast \omega=\Upsilon(\go,C,C)+\Upsilon(\go,C,C,C)+\ldots\,,
\end{equation}
where HS fields are encoded  in two generating functions, the one-form 
\begin{equation}
\label{goc}
\omega(Y,x)=\dr x^\mu \omega_\mu(Y,x)=\sum_{n,m} \dr x^\mu\omega_{\mu\;\alpha_1 \dots \alpha_n,\dot{\alpha}_1\ \dots \dot{\alpha}_m}(x)y^{\alpha_1}\dots y^{\alpha_n}\bar{y}^{\dot{\alpha}_1}\dots \bar{y}^{\dot{\alpha}_m}\; ;\; m+n=2(s-1),
\end{equation}
and zero-form
\begin{equation}\label{Cfield}
C(Y,x)=\sum_{n,m}C_{\alpha_1 \dots \alpha_n,\dot{\alpha}_1\ \dots \dot{\alpha}_m}(x)y^{\alpha_1}\dots y^{\alpha_n}\bar{y}^{\dot{\alpha}_1}\dots \bar{y}^{\dot{\alpha}_m}\; ;\; |m-n|=2s\,
\end{equation}
with two-component indices $\ga\,, \dot \ga=1,2$.
$Y^A=(y^\alpha,\bar{y}^{\dot{\alpha}})$ is $sp(4)$ spinor. Field
components of definite $s$ are associated with  spin-$s$ massless
fields, encoding the original Fronsdal field along with all its
on-shell nontrivial space-time derivatives. In (\ref{HSsketch1}),
(\ref{HSsketch2}) and in the sequel all products of the fields are
wedge products which is implicit. $\dr_x=\dr x^\mu\frac{\p}{\p
x^\mu}$ is space-time De Rham differential.

Star product is defined as follows
\begin{equation}
f(y,\bar{y})\ast g(y,\bar{y})=\int\frac{d^2u d^2v}{(2\pi)^2}\frac{d^2 \bar{u} d^2\bar{v}}{(2\pi)^2}e^{iu_\alpha v^\alpha+i\bar{u}_{\dot{\alpha}} \bar{v}^{\dot{\alpha}}}f(y+u,\bar{y}+\bar{u})g(y+v,\bar{y}+\bar{v}).
\end{equation}

The form of the  vertices on \rhs of \eqref{HSsketch1} and \eqref{HSsketch2}
is determined by the consistency condition with $\dr_x^2=0$. This determines
the vertices up to field redefinitions
\be
\label{fr}
\go' = F(\go,C)\q C' = G(C)\,,
\ee
where $F(\go,C)$ is linear in the one-form $\go$, while both
$F(\go,C)$ and $G(C)$ can be nonlinear in $C$. Indeed, a field
redefinition in the consistent system produces another consistent
system. Since $\go$ and $C$ contain all on-shell nontrivial
derivatives of the Fronsdal fields, non-linear field redefinitions
(\ref{fr}) may contain infinite tails of higher derivatives thus
being non-local though having particular quasi-local form
expandable in power series in terms of components of (\ref{goc})
and (\ref{Cfield}) So, if system (\ref{HSsketch1}),
(\ref{HSsketch2}) is local or spin-local (for more detail see
 \cite{2a2}) in some specific choice
of variables it may lose this property in other variables. Other
way around, if system (\ref{HSsketch1}), (\ref{HSsketch2}) is
non-local in some set of variables, this does not necessarily
imply that there is no other set of variables making the system
spin-local.

Direct computation of vertices consistent with a given locality
requirement from compatibility conditions is technically involved.
 Indeed, deriving HS equations from formal consistency may
result in non-localities due to an inherent natural field
redefinition freedom (see \cite{Vasiliev89} for an earlier
account). To avoid such non-localities one has to impose extra
conditions on field variables that are {\it a priori} unknown.  An
alternative scheme making the derivation of vertices much easier
and setting control on field variables is based on the
generating system of \cite{more}, that has the form
\be
\dr_x W+W*W=0\,,\label{HS1}
\ee
\be
\dr_x S+W*S+S*W=0\,,\label{HS2}
\ee
\be\dr_x B+[W,B]_*=0\,,\label{HS3}
\ee
\be S*S=i(\theta^{A} \theta_{A}+ B*\Gamma) \q \Gamma =\eta \gga +\bar \eta \bar\gga\,,
\label{HS4}\ee
\be
[S,B]_*=0\,,\label{HS5}
\ee
where master fields $W$, $S$ and $B$ depend on space-time
coordinates $x$ and commuting spinor coordinates $Y_A$ and
$Z_A=(z_{\al},\bar z_{\dal})$. In what follows the $x$--dependence
is implicit. In addition there is also a dependence on discrete
involutive Klein elements $K=(k, \bar k)$ such that
(to simplify formulae, in the sequel the star-product symbol $*$
is implicit in the products with Klein elements, \ie
$AK\equiv A*K $)
\be\label{hcom}
\{k,y_{\al}\}=\{k,z_{\al}\}=0\,,\qquad [k,\bar y_{\dal}]=[k,\bar
z_{\dal}]=0\,,\qquad k^2=1\q [k\,,\bar k]=0\,
\ee
and analogously for $\bar k$.

Star product $*$ acts on functions of $Y$ and $Z$  according to
\be\label{star}
(f*g)(Z, Y)=\ff{1}{(2\pi)^4}\int \dr^4 U \dr^4 V f(Z+U; Y+U)g(Z-V;
Y+V)\exp(iU_{A}V^{A})\,,
\ee
where  $sp(4)$ indices $A,B,\dots$ are raised and lowered by the
antisymmetric form $\gep_{AB}=-\gep_{BA}$ as follows
$X^{A}=\gep^{AB}X_{B}$ and $X_A=X^{B}\gep_{BA}$. Master fields are
differential forms with respect to space-time differential
$\dr x^{\nu}$ and auxiliary spinor differential $\theta_A=(\theta_\al,
\bar\theta_{\dal})$ satisfying
\be
\{\theta_{A}\,,\theta_{B}\}=0\q
\{\theta_{\al},k\}=\{\bar\theta_{\dal},\bar k\}=0\,,\qquad
[\theta_{\al},\bar k]=[\bar\theta_{\dal},k]=0\,.
\ee
$B(Z,Y;K|x)$    is a zero-form, while $W(Z,Y;K|x)=\mathrm{d}x^{\mu} W_{\mu}(Z,Y;K|x)$ is a one-form in space-time differential and
$S(Z,Y;K|x)=\theta^{\al} S_{\al}(Z,Y;K|x)+\bar{\theta}^{\dal} \bar{S}_{\dal}(Z,Y;K|x)$ is a one-form in auxiliary spinor differentials. Finally,
$\gga$ and $\bar\gga$ are central two-forms attributed to the Klein operators
\be\label{klein}
\gga=\exp({iz_{\al}y^{\al}})k\theta^{\al} \theta_{\al}\,,\qquad
\bar\gga=\exp({i\bar{z}_{\dal}\bar{y}^{\dal}})\bar
k\bar\theta^{\dal}\bar\theta_{\dal}\,.
\ee
To see that they are central \cite{more}, one should use that $\theta^3 =\bar\theta^3=0$
and that the star-product elements
$\kappa:= \exp({iz_{\al}y^{\al}})$ and $\bar\kappa:=\exp({i\bar{z}_{\dal}\bar{y}^{\dal}})$
have the properties analogous to (\ref{hcom}) with respect to star products with
$f(Y,Z)$ but commute with the Klein operators and differentials $\theta^\ga$ and
$\bar{\theta}^{\dal}$.

\subsection{Perturbation theory}
A proper HS vacuum is the following exact solution of
\eqref{HS1}-\eqref{HS5}
\begin{align}
&B_0=0\,,\label{B0}\\
&S_0=\theta^\al z_{\al}+\bar{\theta}^{\dal}\bar
z_{\dal}\,,\label{S0}\\
&W_0=\go(Y|x)\,,\quad \dr_x\go+\go*\go=0\,.
\end{align}
The flat connection $\go$ can be chosen to describe $AdS_4$.
Since vacuum value of $S_0$ is non-trivial it is going to generate
via star (anti)commutators PDEs in $Z$ for master fields. Indeed, consider equation \eqref{HS5} in the first order
\begin{equation}
[S_0,B_1]_\ast+[S_1,B_0]_\ast=0\,.
\end{equation}
Using  star product \eqref{star} one can check that
\begin{equation}
[Z_A,f(Z,Y;\theta)]_\ast=-2i\frac{\partial}{\partial Z^A}f(Z,Y;\theta).
\end{equation}
It means that $B_1$ field is $Z$ - independent
\begin{equation}
B_1(Z,Y)=C(Y)\,,
\end{equation}
which is the generating function for HS curvatures \eqref{Cfield}. Lower index in $B_1$, shows the order of the expression in the $C$-field. Correspondingly, $B_2$ is of second order in $C$ and so on
\begin{equation}\label{expansionB}
B=C+B_2(C,C)+B_3(C,C,C)+\ldots
\end{equation}
The same rule applies to $S$ and $W$, \ie
\begin{equation}\label{expansionSW}
S=S_0+S_1(C)+S_2(C,C)+\ldots\q W=\omega+W_1(\omega,C)+W_2(\go,C,C)+\ldots
\end{equation}

To solve for $Z$-dependence
of master fields at each perturbation order one has to solve equation of the form
\be\label{steq}
\mathrm{d}_{Z} f(Z;Y;\theta)=J(Z;Y;\theta)\q \dr_Z:=\theta^A \frac{\partial}{\partial Z^A}\,,
\ee
where $J$ originates from the lower-order terms and $f$ is either
$B$, $S$ or $W$. Let us note that it is vacuum solution of the
auxiliary field $S_0$ \eqref{S0} manifesting itself in operator
$\dr_Z$ that allows one to capture the field redefinition
ambiguity as its kernel. Different solutions can be obtained by
one or another contracting homotopy operator $\triangle$
\be
f=\triangle J\,
\ee
resulting from the standard homotopy trick. For an operator $\partial$, that should be also nilpotent,
$\p^2=0$, one considers operator
\begin{equation}
N=\dr_Z \partial+\partial \dr_Z.
\end{equation}
If $N$ is diagonalizable one can introduce the almost inverse
operator $N^\ast$. Since all the $\dr_Z$ - cohomologies are in the
kernel of $N$ one rewrites the solution to \eqref{steq} as
\begin{equation}
f=\partial N^\ast J=\triangle J.
\end{equation}
The simplest choice for $\partial$ is
\begin{equation}
\partial=Z^A\frac{\partial}{\partial \theta^A}\,.
\end{equation}
Then $N$ turns out to be an Euler operator and thus easily invertible
\begin{equation}
N=Z^A\frac{\partial}{\partial Z^A}+\theta^A \frac{\partial}{\partial \theta^A}\q N^\ast J(Z;Y;\theta)=\int_0^1\frac{dt}{t}J(tZ;Y;t\theta).
\end{equation}
This  choice of $\partial$ leads to contracting homotopy operator
\bee\label{oldres}
\hmt_{0}J(Z;Y;\theta)&=&Z^{A}\ff{\p}{\p\theta^{A}}\int_{0}^{1}dt\ff1t
J(t Z; Y; t\theta)\,,
\eee
referred to as the {\it conventional} homotopy operator in
\cite{4a1}.

Using \eqref{HS4} and \eqref{S0} one finds in particular
\begin{equation}
-2i \dr_z S_1^\eta=i\eta C\ast \gamma=i\eta \theta^\alpha \theta_\alpha e^{iz_\alpha y^\alpha}
C(-z\,,\bar y)k\,.
\end{equation}
Using now \eqref{oldres} we  find solution for $S_1^\eta$ in the
form
\begin{equation}\label{S1vvedenie}
S_1^\eta=\eta \theta^\alpha z_\alpha\int_0^1 dt\, t \exp\big\{itz_\alpha y^\alpha\big\}C(-tz,\bar y)k.
\end{equation}

Different choices of homotopy operators represent gauge and field
redefinition ambiguity. A particular class of the so-called
shifted homotopies can be defined by considering $Z^A-Q^A$ instead
of $Z^A$ with some $Z$-independent $Q^A$. Local properties of HS
vertices crucially depend on properties of the chosen homotopy
operators. A class of homotopy operators consistent with locality
requirement based on shifted homotopies at non-trivial interaction
level was proposed in \cite{4a2}.  Let us also note that
another way of fixing the $Z$-dependence based on the so-called
gauge function method is reviewed in \cite{Iazeolla:2020jee}
 (see also \cite{Aros:2019pgj} for applications to various backgrounds).

\subsubsection{Notation}
Let us set up our notation. Derivative with respect to holomorphic
argument of the $C$-fields is denoted as $\partial_{i\,\alpha}$
where index $i$ indicates position of the $C$-field in expression
that contains several $C$'s as seen from left to right. Derivative
with respect to holomorphic argument of $\omega$ - field is
denoted as $\partial_{\go\, \alpha}$.

Whenever arguments of $C$'s or $\go$ are not written explicitly we
assume the {\it exponential} form. This means the following:
suppose one has $\go CCC$. Then it should be understood as
\begin{equation}
\omega(\mathsf{y}_\go,\bar{y})\ust C(\mathsf{y}_1,\bar{y})\ust C(\mathsf{y}_2,\bar{y})\ust C(\mathsf{y}_3,\bar{y})\,,
\end{equation}
{  where $\ust$  denotes the star product with respect to the
 barred variables.}
Derivatives $\partial_\omega$ and $\partial_i$ act as
\begin{equation}
\partial_{\go \, \alpha}=\frac{\partial}{\partial \mathsf{y}_\go^{\alpha}}\q \partial_{i\,\alpha}=\frac{\partial}{\partial \mathsf{y}_i^\alpha}
\end{equation}
followed by all the auxiliary variables set to zero, \ie
\begin{equation}
\mathsf{y}_\go=\mathsf{y}_i=0.
\end{equation}
To make contact with the $p,t$ notation of \cite{4a1} and
\cite{4a2} note that
\begin{equation}
t_\alpha=-i\partial_{\go \, \alpha}\q p_{i\, \alpha}=-i\partial_{i\,\alpha}.
\end{equation}

\section{ Limiting homotopy procedure, subspace $\Hp$ \\and $Z$-dominance lemma}\label{Section3}
The limiting shifted contracting homotopy was introduced in
\cite{4a2} as the generalization of the shifted homotopy
introduced in \cite{4a1}
\begin{multline}
\hmt_{q,\beta}f(z,y\vert \theta)=\\
=\int \frac{d^2u\, d^2 v}{(2\pi)^2}e^{iu_\alpha v^\alpha}\left(z^\alpha+q^\alpha +u^\alpha\right)
\frac{\partial}{\partial \theta^\alpha}\int_0^1 \frac{dt}{t}f(t z -(1-t)(q+u),y+\beta v\vert t \theta)\,,
\end{multline}
where $q^\ga$ is a $z$-independent spinorial shift parameter while
$\gb\in (-\infty , 1)$ is a free parameter. For simplicity we
confine ourselves to the holomorphic sector of undotted spinors
which is of most interest in this paper. (Antiholomorphic sector
of dotted spinors is analysed analogously.)

This operator satisfies the following resolution of identity
\begin{equation}
\label{1}
\dr_z \hmt_{q,\beta}+\hmt_{q,\beta}\dr_z=1-h_{q,\beta}\,,
\end{equation}
where
\begin{equation}\label{projector}
h_{q,\beta}f(z,y\vert \theta)=\int \frac{d^2 u d^2 v}{(2\pi)^2}\,
e^{iu_\alpha v^\alpha}f(-q-u,y+\beta v\vert 0)
\end{equation}
is the projector on $\dr_z$-cohomology.

We say that function $f(z,y)$  of the form
\begin{equation}
\label{class}
f(z,y\vert \theta)=\int_0^1 d\mathcal{T}\, e^{i\mathcal{T}z_\alpha y^\alpha}\phi
\left(\mathcal{T}z,y\vert \mathcal{T} \theta,\mathcal{T}\right)\,
\end{equation}
 belongs to the space $\Hp$ if there exists
 such  real $\varepsilon>0$, that
\begin{equation}\label{limit}
\lim_{\mathcal{T}\rightarrow 0}\mathcal{T}^{1-\varepsilon}\phi(w,u\vert
\theta,\mathcal{T})=0\,.
\end{equation}
Note that the definition of space $\Hp$ is relaxed compared to
that of space $\Sp^{+0}$ of \cite{2a2} because it does not require
any specific behaviour of the $\phi$ at $\mathcal{T}\to1$.
Nevertheless in our calculations sometimes it is convenient to use
 specific form degree relations   of  \cite{2a2} that describe star products
 for the  forms of specific degrees  $p\,,p'$, belonged to spaces $\Sp_p^{0+}$ and $\Sp_{p'}^{+0}$.

There are two main options that appear in the computations below to satisfy \eqref{limit}:
\begin{equation}\label{kernels}
\phi_1(\mathcal{T}z,y\vert \mathcal{T} \theta,
\mathcal{T})=\frac{\mathcal{T}^{\delta_1}}{\mathcal{T}}\widetilde{\phi}_1(\mathcal{T}z,y\vert
\mathcal{T} \theta)\q \phi_2(\mathcal{T} z,y\vert
\mathcal{T}\theta,
\mathcal{T})=\theta(\mathcal{T}-\delta_2)\frac{1}{\mathcal{T}}\widetilde{\phi}_2(\mathcal{T}z,y\vert
\mathcal{T} \theta)
\end{equation}
with some $\delta_{1,2}>0$  and step-function $\theta(x)$. (Note that according to \cite{2a2} the
poles in $\mathcal{T}$ in (\ref{kernels}) are fictitious being
cancelled by the $\mathcal{T}$-dependence of $z$- and
$\theta$-dependent terms in \eq{class}). Functions $\widetilde{\phi}_i(\T
z,y\vert \T \theta)$ are formal power series in variables $\T z$
and $y$ as they are realized as star products of various powers of
$C$'s and $\omega$'s. Functions of the form $\phi_2$ result from the
decomposition $1=\theta(a-\epsilon) +\theta(\epsilon -a)$.

 Space $\Hp$ can be represented as the
direct sum
\begin{equation}
\Hp=\Hp_0 \oplus \Hp_1 \oplus \Hp_2\,,
\end{equation}
where $\Hp_p$ are spanned by the degree-$p$ functions in $\theta$
with kernels that satisfy \eqref{limit}.

Equations of motion (\ref{HSsketch1}), (\ref{HSsketch2}) resulting
from nonlinear system (\ref{HS1})-(\ref{HS5}) have \rhss
independent of $Z^A$ and $\theta^A$ since they belong to the
sector of zero-forms in $\theta$ and are $\dr_Z$-closed as a
consequence of  equations (\ref{HS1})-(\ref{HS5}) resolved at the
previous stages. On the other hand, various terms contributing to
the \rhss of equations (\ref{HSsketch1}), (\ref{HSsketch2}) as a
result of solution of equations (\ref{HS1})-(\ref{HS5}) are of the
form (\ref{class}). In particular, each of these terms is usually
$Z$-dependent. While \rhss of (\ref{HSsketch1}),
(\ref{HSsketch2}) are $Z$-independent as a consequence of
 equations (\ref{HS1})-(\ref{HS5}), the fact that
the sum of all of them is $Z$-independent is not obvious,
demanding an appropriate partial integrations over homotopy
parameters that appear at various stages of the order-by-order
analysis of nonlinear HS equations. After all, functions
(\ref{class}) can be $Z$-independent only if they have a
distributional measure supported at $\mathcal{T}=0$, \ie after
appropriate partial integrations the measure contains a factor of
$\delta(\mathcal{T})$. Such a measure has dimension $-1$ in
$\mathcal{T}$. If a function contains an additional factor of
$\mathcal{T}^\gvep$, it cannot contribute to the  $Z$-independent
answer. This just means that functions of the class $\Hp_0$ cannot
contribute to the $Z$-independent equations  (\ref{HSsketch1}),
(\ref{HSsketch2}). This is the content of $Z$-dominance Lemma of
\cite{2a1}: any terms in $\phi(w,u\vert \theta, \mathcal{T}) $
dominated by a positive power of $\mathcal T$ do not contribute to
the dynamical equations (\ref{HSsketch1}), (\ref{HSsketch2}).
Application of this fact to locality is straightforward once it is
shown that all terms containing
infinite towers of higher derivatives in the vertices of interest
belong to $\Hp_0$ and, therefore, do not contribute to HS
equations (\ref{HSsketch2}). This is what is shown in this paper.

A related fact is that, as shown in \cite{4a2}, the space $\Hp$ exhibits special properties under the action of the
limiting shifted  homotopy $\hmt_{q,\beta}$ at $\gb\to-\infty$   leading to
local HS interactions. Namely, it maps  $\Hp_1$ to $\Hp_0$  \cite{2a2},
\begin{equation}\label{classSave}
\lim_{\beta \rightarrow -\infty} \hmt_{q,\beta} f_1(z,y\vert \theta)=f_0(z,y\vert 0)
\q \forall f_1\in \Hp_1\q f_0 \in \Hp_0\,.
\end{equation}
Since elements of $\Hp_0$ do not contribute to $Z$-independent
physical vertex by $Z$-dominance lemma, this property allows us to discard all terms from
$\Hp$ in the analysis of the $\go C^3$ vertex in equation
(\ref{HSsketch2}).

More in detail,
to prove that the vertex is spin-local we find
it most efficient to represent the \rhs
of equations in the $\dr_Z$--exact spin-local form modulo terms from $\Hp$ that do not contribute to
the final vertex by $Z$-dominance lemma.
In other words, to solve equation (\ref{steq}) we represent $J$ on its \rhs  in the form
\be
\label{ex}
J=\dr_Z \widehat{f} +J^+\q J^+\in \Hp\,,
\ee
where $\widehat{f}$ is spin-local. Since $\dr_Z J=0$ and $\dr_Z^2=0$, we conclude that
\be
\dr_Z J^+=0\,.
\ee
We now note that for the sum of two $\dr_Z$-closed terms one can
use independent contracting homotopies for each of the summands.
Hence, we can write for $f$ on the \lhs of (\ref{steq})
\be
\label{fex}
f=\widehat{f} + \lim_{\gb\to-\infty}\hmt_{q,\gb} J^+
\ee
with some $q$. Since $J^+\in \Hp$, with this definition the contribution of $J^+$
will give zero to the dynamical field equations for any $q$ while $\widehat{f}$ will give a
spin-local contribution. Details of the derivation of the decomposition (\ref{ex}) in the $B$ and $W$ sectors are given in Appendices B and C, respectively.

 To put it differently, to eliminate $Z$-dependence of a
seemingly $Z$-dependent expression we manage  to show that the
vertices $\Upsilon^{\eta\eta}(\go,C,C,C)$ and
$\Upsilon^{\bar{\eta}\bar{\eta}}(\go,C,C,C)$ are spin-local modulo
terms from $\Hp$ that vanish by $Z$-dominance lemma, \ie
\be\label{schemUps}\qquad \Upsilon(Y)= \widehat{\Upsilon}(Z,Y)+ \Upsilon_+(Z,Y)\q \Upsilon_+\in   \Hp\,,\ee
where $\widehat{\Upsilon}$ is spin-local but  $Z$-dependent expression.
Now we observe that since the physical  vertex $ \Upsilon(Y)$ is $\theta$, $Z$-independent it can be written in
the form
\be
 \Upsilon(Y)= \widehat{\Upsilon}(Z,Y)+ \Upsilon_+(Z,Y) = h_{q,\gb} (\widehat{\Upsilon}(Z,Y)+ \Upsilon_+(Z,Y))
\ee
where $ h_{q,\gb}$ is the cohomology projector with any shift parameters $q$ and $\gb$ (recall that,
as is obvious from (\ref{projector}), cohomology
projectors leave $\theta, Z$-independent functions invariant \cite{4a2}). Taking the limit
$\gb\to-\infty$ we find that, for any  $q$,  components of
 $\widehat{\Upsilon}(Z,Y)$ can  contribute  to the resulting vertex $\Upsilon(Y)$ while $\Upsilon_+(Z,Y)$ cannot
 since  the limiting projector $h_{q,-\infty}$ \eqref{projector} acts trivially on
$\Hp_0$ \cite{2a2},
\begin{equation}
\label{coh}
\lim_{\beta \rightarrow -\infty} h_{q,\beta} f_0(Z,Y\vert 0)=0\q f_0 \in \Hp_0\,.
\end{equation}
This implies that
 \be h_{q,-\infty}(\widehat{\Upsilon}(Z,Y)+ \Upsilon_+(Z,Y))=h_{q,-\infty}(\widehat{\Upsilon}(Z,Y) )\,.
\ee
This formula provides an alternative interpretation of $Z$-dominance Lemma stating that
elements of $\Hp_0$ do not contribute to the physical $Z$-independent vertices though
practically, its application with any $q$ may not be useful since
 in most cases the result has a seemingly non-local form.

On the other hand, by $Z$-dominance lemma, a spin-local vertex $\widehat{\Upsilon}(Z,Y)$ must be
decomposable into a sum
 \be
 \label{ll}
  \widehat{\Upsilon}(Z,Y)=\Upsilon^{loc}(Y) + \Upsilon^{loc}_+(Z,Y) \q \Upsilon^{loc}_+\in   \Hp
\ee
with a spin-local $Z$-independent $\Upsilon^{loc}(Y)$.
Hence one can conclude that $\Upsilon^{loc}_+(Z,Y)+\Upsilon
_+(Z,Y)=0$ which fact  is not at all manifest being a consequence
of Schouten identity and
various relations between integrals  over homotopy parameters.

Of course, once $\Upsilon^{loc}(Y)$ is found, the application of
$h_{q,-\infty}$ with any $q$ to the \rhs of \eq{schemUps} gives a
spin-local vertex $\Upsilon (Y)=\Upsilon^{loc}(Y)$. It should be
stressed again however that, though one can formally obtain the final
result via application of any cohomology projector, it will be
spin-local, but not manifestly spin-local, containing  non-local
contributions that will all cancel by virtue of partial
integrations and Schouten identity used in the derivation of the
decomposition (\ref{ll}) which is practically easier to find. As
will be demonstrated  in the
forthcoming paper \cite{GelKor}, being technically involved,
this approach makes it possible to compute explicit form of the
physical spin-local vertices.

\section{Final results}\label{RES}
\subsection{General structure of equations}
Dynamical equations up to the third order in the zero-forms $C$ can be schematically
put into the form
\begin{equation}\label{EOMsch}
\dr_x C+[\go,C]_\ast=\Upsilon^\eta(\go,C,C)+\Upsilon^{\bar{\eta}}(\go,C,C)+\Upsilon^{\eta\eta}(\go,C,C,C)+\Upsilon^{\bar{\eta}\bar{\eta}}(\go,C,C,C)+\Upsilon^{\eta\bar{\eta}}(\go,C,C,C)+\ldots
\end{equation}
The vertex $\Upsilon^{\eta\eta}(\go,C,C,C)$ resulting
 from system \eqref{HS1}-\eqref{HS5} has the form
\begin{equation}\label{rhs}
\Upsilon^{\eta\eta}(\go,C,C,C)=-\dr_x B_3^{\eta\eta}-\dr_x B_2^{\eta}-
[\go,B_3^{\eta\eta}]_\ast-[W_1^\eta,B_2^\eta]_\ast-[W_2^{\eta\eta},C]_\ast\,.
\end{equation}
Here $W^\eta_1,W_2^{\eta\eta}$ and $B_2^\eta, B_3^{\eta\eta}$ are master fields of the corresponding orders from expansions \eqref{expansionB}, \eqref{expansionSW} which are to be obtained from the generating system via solving equation of the type \eqref{steq}. Each term on the \rhs of this equation depends both on $Y$ and on $Z$. These vertices can be decomposed into two parts
\begin{equation}\label{upsz}
\Upsilon^{\eta\eta}(\go,C,C,C)=\widehat{\Upsilon}^{\eta\eta}(\go,C,C,C)+
{\Upsilon}_+^{\eta\eta}(\go,C,C,C)\q
{\Upsilon}_+^{\eta\eta}(\go,C,C,C)\in \Hp_0\,,
\end{equation}
where unlike the whole $\Upsilon^{\eta\eta}(\go,C,C,C)$ its two
contributions on the right of \eqref{upsz} can be $z$--dependent.

In this paper we compute the
$\widehat{\Upsilon}^{\eta\eta}(\go,C,C,C)$ part of the vertices.
This part turns out to  be free from
contractions between holomorphic variables of the $C$-fields because such terms
belong to $\Hp_0$. Consistency of
equations \eqref{HS1}-\eqref{HS5}
guarantees that $\Upsilon^{\eta\eta}(\go,C,C,C)$ is $Z$-independent
and, according to $Z$-dominance Lemma, it can be realized only as $\delta(\T)$ in the kernel.
Hence $Z$-independent expression for $\Upsilon^{\eta\eta}(\go,C,C,C)$ must be free from infinite tower of contractions
 between holomorphic variables  which implies spin-locality of the resulting HS equations.

In this section  we present final expression for $\widehat{\Upsilon}^{\eta\eta}(\go,C,C,C)$
\begin{equation}
\label{V}
\widehat{\Upsilon}^{\eta\eta}(\go,C,C,C)=\widehat{\Upsilon}^{\eta\eta}_{\go CCC}
+\widehat{\Upsilon}_{C\go CC}^{\eta\eta}+\widehat{\Upsilon}_{CC\go C}^{\eta\eta}
+\widehat{\Upsilon}_{CCC\go}^{\eta\eta}
\end{equation}
obtained from the generating system \eqref{HS1}-\eqref{HS5} using the perturbation
scheme up to the third order in $C$-field. Details of their derivation are presented
in Appendices B and C.

The vertices in (\ref{V}) are composed  from the following terms

\begin{equation}\label{goCCC}
\widehat{\Upsilon}^{\eta\eta}_{\go CCC}\approx-\dr_x \widehat{B}_3^{\eta\eta}\Big|_{\go CCC}
- \omega\ast \widehat{B}_3^{\eta\eta}-\dr_x B_2^{\eta\, loc}\Big|_{\go CCC}-W^{\eta}_{1\, \go C}\ast
B_2^{\eta\, loc}-\widehat{W}_{2\, \go CC}^{\eta\eta}\ast C,
\end{equation}
\begin{equation}
\widehat{\Upsilon}_{C\go CC}^{\eta\eta}\approx-\dr_x \widehat{B}_3^{\eta\eta}\Big|_{C\go CC}
-\dr_x B_2^{\eta\, loc}\Big|_{C\go CC}-W^{\eta}_{1\, C\go}\ast B_2^{\eta\, loc}
-\widehat{W}_{2\, C\go C}^{\eta\eta}\ast C+C\ast \widehat{W}_{2\, \go CC}^{\eta\eta},
\end{equation}
\begin{equation}
\widehat{\Upsilon}^{\eta\eta}_{CC\go C}\approx-\dr_x \widehat{B}_3^{\eta\eta}\Big|_{CC\go C}-\dr_x
B_2^{\eta\, loc}\Big|_{CC\go C}+B_2^{\eta\, loc}\ast W^\eta_{1\, \go C}
-W^{\eta\eta}_{2\, CC\go}\ast C+C\ast W^{\eta\eta}_{2\, C\go C},
\end{equation}
\begin{equation}\label{CCCgo}
\widehat{\Upsilon}_{CCC\go}^{\eta\eta}\approx-\dr_x
\widehat{B}_3^{\eta\eta}\Big|_{CCC\go}+\widehat{B}_3^{\eta\eta}\ast
\go-\dr_x B_2^{\eta\, loc}\Big|_{CCC\go}+B_2^{\eta\, loc}\ast
W_{1\, C\go}^{\eta}+C\ast \widehat{W}_{2\, CC\go}^{\eta\eta}.
\end{equation}

The expression  for $B_2^{\eta\, loc}$
has the form  \cite{Vas}
\begin{multline}\label{B2loc}
B_2^{\eta\, loc}=\frac{\eta}{2}\int d^3\tau_+ \big[\delta^\prime (1-\sum_{i=1}^{3}\tau_i)
-iz_\alpha y^\alpha \delta(1-\sum_{i=1}^{3}\tau_i)\big]\exp({i\tau_1\,z_\alpha y^\alpha
+i\tau_1 \partial_{1\alpha}\partial_2 {}^\alpha})\times\\
C(-\tau_1 z+\tau_2 y,\bar{y})\ust C(-\tau_1 z-\tau_3 y,\bar{y})k\,,
\end{multline}
where we use a short-hand notation
\be
d^3\tau_+:= d\tau_1 d\tau_2 d\tau_3 \theta(\tau_1) \theta(\tau_2) \theta(\tau_3)\,.
\ee
 Whenever this notation is used there is always a delta-function in a corresponding expression  $\delta(1-\tau_1-\tau_2-\tau_3)$ that bounds the $\tau_i$ variables from above.
 Note that
 $B_2^{\eta loc}$  is a sum of $B_2^\eta$ obtained in \cite{4a1} and
local cohomology (\ie $Z$-independent) shift $\delta B_2^\eta$
\bee\label{redef}
B_2^{\eta loc}&=&B_2^\eta+\delta B_2^\eta\q \\ \nn
\delta B_2^\eta&=&\frac{\eta}{2}\int d^2\tau_+
\delta(1-\tau_1-\tau_2)C(\tau_1 y,\bar{y})\ust  C(-\tau_2y,\bar{y})k.
\eee

The expressions   $W_{1\, \go C}^{\eta}$ and  $W_{1\, C\go}^\eta$ were obtained  in \cite{4a1}, having the form
\begin{multline}\label{W1goCeta}
W_{1\, \omega C}^\eta=-\frac{\eta}{2}\int_0^1 d\tau_1 \int_0^1 d\sigma\, (1-\tau_1)
\left(z^\alpha \partial_{\omega \alpha}\right)\exp\Big\{ i\tau_1\, z_\alpha y^\alpha
+i(1-(1-\tau_1)\sigma)\partial_{\omega\alpha}\partial_1 {}^\alpha\Big\}\times \\
\omega(-\tau_1 z+(1-\tau_1)\sigma y,\bar{y})\ust C(-\tau_1 z,\bar{y})k\,,
\end{multline}
\begin{multline}\label{W1Cgoeta}
W_{1\,C\omega}^\eta=-\frac{\eta}{2}\int_0^1 d\tau_1 \int_0^1 d\sigma\,
 (1-\tau_1)\left(z^\alpha \partial_{\omega \alpha}\right)\exp\Big\{i\tau_1 z_\alpha y^\alpha
 +i(1-(1-\tau_1)\sigma)\partial_{1\alpha}\partial_{\go}{}^\alpha\Big\}\times\\
 C(-\tau_1 z,\bar{y})\ust \go(-\tau_1 z-(1-\tau_1)\sigma y,\bar{y})k\,.
\end{multline}

Note that the terms with $\dr_x$ contribute to the third order via the second-order
contribution to $\dr_x C$
\be\label{C2EOM}
\dr_x C = C\ast \omega -\omega \ast C+\Upsilon^\eta_{\go CC}+\delta \Upsilon^\eta_{\go CC}
+\Upsilon^{\eta}_{C\go C}+\delta\Upsilon^{\eta}_{C\go C}+\Upsilon^{\eta}_{CC\go}
+\delta \Upsilon^{\eta}_{CC\go}\,.
\ee
Here $\Upsilon^\eta_{\go CC}, \Upsilon^{\eta}_{C\go C}$ and
$\Upsilon^{\eta}_{CC\go}$ are vertices obtained in \cite{4a1} and
$\delta\Upsilon^\eta_{\go CC}, \delta\Upsilon^{\eta}_{C\go
C},\delta\Upsilon^{\eta}_{CC\go}$  result from the  local field
redefinition of  $B_2^\eta$
 \eq{redef} giving
 \begin{multline}\label{VgoCC}
\Upsilon_{\go CC}^\eta +\delta \Upsilon_{\go CC}^\eta=-\frac{\eta}{2}\int d^3\tau_+\,
\delta\left(1-\sum_{i=1}^3 \tau_i\right)\Big(y^\alpha \partial_{\omega \alpha}\Big)
\exp\Big\{i(1-\tau_2)\partial_{\omega \alpha}\partial_1^\alpha
-i\tau_2 \partial_{\omega \alpha}\partial_2^\alpha\Big\}\times\\
\times \omega\big((1-\tau_3)y,\bar{y}\big)\ust C\big(\tau_1 y,\bar{y}\big)\ust C\big((\tau_1-1)y,\bar{y}\big)k,
\end{multline}
\begin{multline}\label{VCCgo}
\Upsilon_{CC\go}^\eta+\delta \Upsilon_{CC\go}^\eta=-\frac{\eta}{2}\int d^3\tau_+
\,\delta\left(1-\sum_{i=1}^3 \tau_i\right)\Big(y^\alpha \partial_{\omega \alpha}\Big)
\exp\Big\{i(1-\tau_2)\partial_{2 \alpha}\partial_{\go}^\alpha
-i\tau_2\partial_{1\alpha} \partial_{\go}^\alpha\Big\}
\times\\\times C\big((1-\tau_1)y,\bar{y}\big)\ust C\big(-\tau_1 y,\bar{y}\big)
\ust \omega\big((\tau_3-1)y,\bar{y}\big)k,
\end{multline}
\begin{multline}\label{VCgoC}
\Upsilon^\eta_{C\go C}+\delta \Upsilon^\eta_{C\go C}=-\frac{\eta}{2}\int d^3 \tau_+\,
\delta\left(1-\sum_{i=1}^3 \tau_i\right)\Big(y^\alpha \partial_{\omega \alpha}\Big)
\exp\Big\{i\tau_2 \partial_{1\alpha}\partial_{\go}^\alpha+i(1-\tau_2)\partial_{\omega \alpha}\partial_2^\alpha
\Big\}
\times\\
\times C\big(\tau_3 y,\bar{y}\big)\ust \omega\big(-\tau_1 y,\bar{y}\big)\ust C\big((\tau_3-1)y,\bar{y}\big)k\\
-\frac{\eta}{2}\int d^3 \tau_+\, \delta\left(1-\sum_{i=1}^3 \tau_i\right)
\Big(y^\alpha \partial_{\omega \alpha}\Big)\exp\Big\{i(1-\tau_2)\partial_{1\alpha}\partial_{\go}^\alpha
+i\tau_2 \partial_{\omega \alpha}\partial_2^\alpha\Big\}\times\\
\times C\big((1-\tau_3)y,\bar{y}\big)\ust \omega\big(\tau_1 y,\bar{y}\big)\ust C\big(-\tau_3 y,\bar{y}\big)k.
\end{multline}

\subsection{The fields}

The expressions for $\widehat{B}^{\eta\eta}_3$ and $\widehat{W}^{\eta\eta}_2$ derived in Appendices B and С are

\begin{multline}\label{B3final}
\widehat{B}_{3}^{\eta\eta}=-\frac{\eta^2}{4} \int_0^1 d\mathcal{T}\, \mathcal{T}  \int d^3 \rho_+ \delta\left(1-\sum_{i=1}^3 \rho_i\right)   \int_0^1 d\xi\, \frac{\rho_1\, (z_\alpha y^\alpha)^2 }{(\rho_1+\rho_2)(\rho_1+\rho_3)}\times\\
\exp\Big\{i\mathcal{T}\, z_\alpha y^\alpha+\mathcal{T} z^\alpha\Big(-(\rho_1+\rho_3)\partial_{1\alpha}+(\rho_2-\rho_3)\partial_{2\alpha}+(\rho_1+\rho_2)\partial_{3\alpha}\Big)\\
+(1-\xi)y^\alpha\left(\frac{\rho_1}{\rho_1+\rho_2}\partial_{1\alpha}-\frac{\rho_2}{\rho_1+\rho_2}\partial_{2\alpha}\right)+\xi\, y^\alpha\left(\frac{\rho_1}{\rho_1+\rho_3}\partial_{3\alpha}-\frac{\rho_3}{\rho_1+\rho_3}\partial_{2\alpha}\right)\Big\}CCC\,,
\end{multline}
\begin{multline}\label{W2CCgofinal}
\widehat{W}_{2\, CC\go}^{\eta\eta}=\frac{\eta^2}{4}\int_0^1 d\mathcal{T}\, \mathcal{T}
\int d^4 \rho_+\, \delta\left(1-\sum_{i=1}^4 \rho_i\right)
\frac{\rho_1\left(z^\gamma \partial_{\go \gamma}\right)^2}{(\rho_1+\rho_2)(\rho_3+\rho_4)}\times\\
\times\exp\Big\{i\mathcal{T}z_\alpha y^\alpha+\mathcal{T}z^\alpha
\Big(-(\rho_1+\rho_2)\partial_{1 \alpha}+(\rho_3+\rho_4)\partial_{2 \alpha}+(1-\rho_2)\partial_{\go \alpha}\Big)\\
+\frac{\rho_1\rho_3}{(\rho_1+\rho_2)(\rho_3+\rho_4)}y^\alpha \partial_{\go \alpha}
+i\left(\frac{(1-\rho_4)\rho_2}{\rho_1+\rho_2}+\rho_4\right)\partial_{2 \alpha}\partial_{\go} {}^\alpha
-i\frac{\rho_1\rho_4}{\rho_3+\rho_4}\partial_{1 \alpha}\partial_{\go} {}^\alpha\Big\}CC\go,
\end{multline}
\begin{multline}\label{W2goCCfinal}
\widehat{W}_{2\, \go CC}^{\eta\eta}=\frac{\eta^2}{4}\int_0^1 d\mathcal{T}\,\T
\int d^4\rho_+\, \delta\left(1-\sum_{i=1}^4 \rho_i\right)
\frac{\rho_1 \left(z^\gamma \partial_{\go \gamma}\right)^2}{(\rho_1+\rho_2)(\rho_3+\rho_4)}\times\\
\times \exp\Big\{i\mathcal{T}z_\alpha y^\alpha+\mathcal{T}z^\alpha\Big((1-\rho_2)\partial_{\go\alpha}-(\rho_3+\rho_4)\partial_{1\alpha}+(\rho_1+\rho_2)\partial_{2 \alpha}\Big)\\
+\frac{\rho_1\rho_3}{(\rho_1+\rho_2)(\rho_3+\rho_4)}y^\alpha \partial_{\go \alpha}
+i\left(\frac{(1-\rho_4)\rho_2}{\rho_1+\rho_2}+\rho_4\right)\partial_{\go \alpha}\partial_1 {}^\alpha
-i\frac{\rho_4\rho_1}{\rho_3+\rho_4}\partial_{\go \alpha}\partial_2 {}^\alpha\Big\}\go CC,
\end{multline}
\begin{multline}\label{W2CgoCfinal}
\widehat{W}_{2\, C\go C}^{\eta \eta}=-\frac{\eta^2}{2}\int_0^1 d\mathcal{T}\, \mathcal{T}
 \int d^4\rho_+\, \delta\left(1-\sum_{i=1}^4 \rho_i\right)
 \frac{(\rho_1+\rho_3)\left(z^\gamma\partial_{\go \gamma}\right)^2}{(\rho_1+\rho_2)(\rho_3+\rho_4)}\times\\
\times \exp\Big\{i\mathcal{T}z_\alpha y^\alpha+\mathcal{T}z^\alpha
\Big(-(\rho_3+\rho_4)\partial_{1 \alpha}+(\rho_1-\rho_3)\partial_{\go \alpha}
+(\rho_1+\rho_2)\partial_{2 \alpha}\Big)
\\-\frac{\rho_3\rho_1}{(\rho_1+\rho_2)(\rho_3+\rho_4)}y^\alpha \partial_{\go\alpha}
+i\frac{\rho_4(1-\rho_2)}{\rho_3+\rho_4}\partial_{\go \alpha}\partial_{2} {}^\alpha
+i\frac{\rho_2(1-\rho_4)}{\rho_1+\rho_2}\partial_{1 \alpha}\partial_\go {}^\alpha\Big\}C\go C.
\end{multline}
From now on we skip antiholomorphic
 (barred) variables for brevity.  More precisely,
$CCC$ on the \rhs of \eq{B3final}
is to be understood as $C(\mathsf{y}_1, \by)\ust C (\mathsf{y}_2, \by)\ust C(\mathsf{y}_3, \by)
\big|_{\mathsf{y}_i=0}$,
$\go CC$ on the \rhs of \eq{W2goCCfinal}
as $\go(\mathsf{y}_\go, \by)\ust C(\mathsf{y}_1, \by)\ust C (\mathsf{y}_2, \by)
\big|_{\mathsf{y}_\go=\mathsf{y}_i=0}$ {\it etc}.

Expressions \eqref{B3final}-\eqref{W2CgoCfinal} are spin-local
because the exponential factors in all of them are free from terms
$\p_{i\ga}\p_j^\ga$ describing contractions between higher
components of the zero-forms $C(Y)$ bringing higher-derivative
vertices for fields of particular spins. So are the terms induced
by these expressions in vertices \eqref{goCCC}-\eqref{CCCgo}.
Indeed, differentiating $\widehat{B}_3^{\eta\eta}$ one should use
only first-order part from \rhs of \eqref{C2EOM} which does not
bring contractions between $C$-fields, similarly star product with
$\omega$ does not bring contractions due to \eqref{YL},
\eqref{YR}. On the other hand, though star product of
$\widehat{W}_2^{\eta\eta}$ with $C$ does bring contractions
between the fields $C$, all of them result from the $z$-dependent
terms in the exponentials (\ref{W2CCgofinal})-(\ref{W2CgoCfinal})
that carry at least one power of $\mathcal{T}$. Such terms contain
an additional factor of $\T$ in front of the contraction terms
$\p_{i\ga}\p_j^\ga$ thus belonging to $\Hp$. Hence all the
contributions to the vertex (\ref{rhs}) induced from
$B_3^{\eta\eta}$ and $W_2^{\eta\eta}$ are spin-local modulo terms
in $\Hp$.

\subsection{Equations}

\subsubsection{$B_3$ driven terms}
Direct computation of the $B_3$ induced terms using \eqref{C2EOM},
\eqref{B3final} and \eqref{YL}, \eqref{YR} gives
\begin{multline}
\dr_x \widehat{B}_3^{\eta\eta}\big|_{\go
CCC}\approx\frac{\eta^2}{4} \int_0^1 d\mathcal{T}\, \mathcal{T}
\int d^3 \rho_+ \delta\left(1-\sum_{i=1}^3 \rho_i\right) \int_0^1
d\xi\, \frac{\rho_1\, (z_\alpha y^\alpha)^2
e^{i\mathcal{T}\, z_\alpha y^\alpha} }{(\rho_1+\rho_2)(\rho_1+\rho_3)}\times\\
\times\exp\Big\{\mathcal{T} z^\alpha
\Big(-(\rho_1+\rho_3)(\partial_{\go\alpha}+\partial_{1\alpha})+(\rho_2-\rho_3)\partial_{2\alpha}
+(\rho_1+\rho_2)\partial_{3\alpha}\Big)+i\partial_{\go \alpha}\partial_1 {}^\alpha\\
+(1-\xi)y^\alpha\left(\frac{\rho_1}{\rho_1+\rho_2}(\partial_{\go \alpha}+\partial_{1\alpha})-\frac{\rho_2}{\rho_1+\rho_2}\partial_{2\alpha}\right)+\xi\, y^\alpha\left(\frac{\rho_1}{\rho_1+\rho_3}\partial_{3\alpha}-\frac{\rho_3}{\rho_1+\rho_3}\partial_{2\alpha}\right)\Big\}\go CCC,
\end{multline}
\begin{multline}
\dr_x \widehat{B}_3^{\eta\eta}\big|_{C\go CC}\approx-\frac{\eta^2}{4} \int_0^1 d\mathcal{T}\, \mathcal{T}  \int d^3 \rho_+ \delta\left(1-\sum_{i=1}^3 \rho_i\right)   \int_0^1 d\xi\, \frac{\rho_1\, (z_\alpha y^\alpha)^2 e^{i\mathcal{T}\, z_\alpha y^\alpha} }{(\rho_1+\rho_2)(\rho_1+\rho_3)}\times\\
\times\exp\Big\{\mathcal{T} z^\alpha\Big(-(\rho_1+\rho_3)(\partial_{\go\alpha}+\partial_{1\alpha})+(\rho_2-\rho_3)\partial_{2\alpha}+(\rho_1+\rho_2)\partial_{3\alpha}\Big)+i\partial_{1 \alpha}\partial_\go {}^\alpha\\
+(1-\xi)y^\alpha\left(\frac{\rho_1}{\rho_1+\rho_2}(\partial_{\go \alpha}+\partial_{1\alpha})-\frac{\rho_2}{\rho_1+\rho_2}\partial_{2\alpha}\right)+\xi\, y^\alpha\left(\frac{\rho_1}{\rho_1+\rho_3}\partial_{3\alpha}-\frac{\rho_3}{\rho_1+\rho_3}\partial_{2\alpha}\right)\Big\} C\go CC\\
+\frac{\eta^2}{4} \int_0^1 d\mathcal{T}\, \mathcal{T}  \int d^3 \rho_+ \delta\left(1-\sum_{i=1}^3 \rho_i\right)   \int_0^1 d\xi\, \frac{\rho_1\, (z_\alpha y^\alpha)^2 e^{i\mathcal{T}\, z_\alpha y^\alpha} }{(\rho_1+\rho_2)(\rho_1+\rho_3)}\times\\
\times\exp\Big\{\mathcal{T} z^\alpha\Big(-(\rho_1+\rho_3)\partial_{1\alpha}+(\rho_2-\rho_3)(\partial_{\go\alpha}+\partial_{2\alpha})+(\rho_1+\rho_2)\partial_{3\alpha}\Big)+i\partial_{\go \alpha}\partial_2 {}^\alpha\\
+(1-\xi)y^\alpha\left(\frac{\rho_1}{\rho_1+\rho_2}\partial_{1\alpha}-\frac{\rho_2}{\rho_1+\rho_2}(\partial_{\go\alpha}+\partial_{2\alpha})\right)\\
+\xi\, y^\alpha\left(\frac{\rho_1}{\rho_1+\rho_3}\partial_{3\alpha}-\frac{\rho_3}{\rho_1+\rho_3}(\partial_{\go \alpha}+\partial_{2\alpha})\right)\Big\} C\go CC,
\end{multline}

\begin{multline}
\dr_x \widehat{B}_3^{\eta\eta}\big|_{CC\go C}\approx-\frac{\eta^2}{4} \int_0^1 d\mathcal{T}\, \mathcal{T}  \int d^3 \rho_+ \delta\left(1-\sum_{i=1}^3 \rho_i\right)   \int_0^1 d\xi\, \frac{\rho_1\, (z_\alpha y^\alpha)^2 e^{i\mathcal{T}\, z_\alpha y^\alpha} }{(\rho_1+\rho_2)(\rho_1+\rho_3)}\times\\
\times\exp\Big\{\mathcal{T} z^\alpha\Big(-(\rho_1+\rho_3)\partial_{1\alpha})+(\rho_2-\rho_3)(\partial_{\go \alpha}+\partial_{2\alpha})+(\rho_1+\rho_2)\partial_{3\alpha}\Big)+i\partial_{2 \alpha}\partial_\go {}^\alpha\\
+(1-\xi)y^\alpha\left(\frac{\rho_1}{\rho_1+\rho_2}\partial_{1\alpha}-\frac{\rho_2}{\rho_1+\rho_2}(\partial_{\go \alpha}+\partial_{2\alpha})\right)+\xi\, y^\alpha\left(\frac{\rho_1}{\rho_1+\rho_3}\partial_{3\alpha}-\frac{\rho_3}{\rho_1+\rho_3}(\partial_{\go \alpha}+\partial_{2\alpha})\right)\Big\} CC\go C\\
+\frac{\eta^2}{4} \int_0^1 d\mathcal{T}\, \mathcal{T}  \int d^3 \rho_+ \delta\left(1-\sum_{i=1}^3 \rho_i\right)   \int_0^1 d\xi\, \frac{\rho_1\, (z_\alpha y^\alpha)^2 e^{i\mathcal{T}\, z_\alpha y^\alpha} }{(\rho_1+\rho_2)(\rho_1+\rho_3)}\times\\
\times\exp\Big\{\mathcal{T} z^\alpha\Big(-(\rho_1+\rho_3)\partial_{1\alpha}+(\rho_2-\rho_3)\partial_{2\alpha}+(\rho_1+\rho_2)(\partial_{\go\alpha}+\partial_{3\alpha})\Big)+i\partial_{\go \alpha}\partial_3 {}^\alpha\\
+(1-\xi)y^\alpha\left(\frac{\rho_1}{\rho_1+\rho_2}\partial_{1\alpha}-\frac{\rho_2}{\rho_1+\rho_2}\partial_{2\alpha}\right)
+\xi\, y^\alpha\left(\frac{\rho_1}{\rho_1+\rho_3}(\partial_{\go \alpha}+\partial_{3\alpha})-\frac{\rho_3}{\rho_1+\rho_3}\partial_{2\alpha}\right)\Big\} CC\go C,
\end{multline}

\begin{multline}
\dr_x \widehat{B}_3^{\eta\eta}\big|_{ CCC\go}\approx-\frac{\eta^2}{4} \int_0^1 d\mathcal{T}\, \mathcal{T}  \int d^3 \rho_+ \delta\left(1-\sum_{i=1}^3 \rho_i\right)   \int_0^1 d\xi\, \frac{\rho_1\, (z_\alpha y^\alpha)^2 e^{i\mathcal{T}\, z_\alpha y^\alpha} }{(\rho_1+\rho_2)(\rho_1+\rho_3)}\times\\
\times\exp\Big\{\mathcal{T} z^\alpha\Big(-(\rho_1+\rho_3)\partial_{1\alpha}+(\rho_2-\rho_3)\partial_{2\alpha}+(\rho_1+\rho_2)(\partial_{\go \alpha}+\partial_{3\alpha})\Big)+i\partial_{3 \alpha}\partial_\go {}^\alpha\\
+(1-\xi)y^\alpha\left(\frac{\rho_1}{\rho_1+\rho_2}\partial_{1\alpha}-\frac{\rho_2}{\rho_1+\rho_2}\partial_{2\alpha}\right)+\xi\, y^\alpha\left(\frac{\rho_1}{\rho_1+\rho_3}(\partial_{\go \alpha}+\partial_{3\alpha})-\frac{\rho_3}{\rho_1+\rho_3}\partial_{2\alpha}\right)\Big\}CCC\go.
\end{multline}
\begin{multline}
\omega\ast \widehat{B}_3^{\eta\eta}\approx-\frac{\eta^2}{4} \int_0^1 d\mathcal{T}\, \mathcal{T}  \int d^3 \rho_+ \delta\left(1-\sum_{i=1}^3 \rho_i\right)   \int_0^1 d\xi\, \frac{\rho_1\, \left[z_\alpha\left(y^\alpha-i\partial_{\go} {}^\alpha\right)\right]^2 e^{i\mathcal{T}\, z_\alpha y^\alpha} }{(\rho_1+\rho_2)(\rho_1+\rho_3)}\times\\
\times\exp\Big\{\mathcal{T} z^\alpha\Big(-\partial_{\go \alpha}-(\rho_1+\rho_3)\partial_{1\alpha}+(\rho_2-\rho_3)\partial_{2\alpha}+(\rho_1+\rho_2)\partial_{3\alpha}\Big)+y^\alpha\partial_{\go \alpha}\\
+(1-\xi)y^\alpha\left(\frac{\rho_1}{\rho_1+\rho_2}\partial_{1\alpha}-\frac{\rho_2}{\rho_1+\rho_2}\partial_{2\alpha}\right)+\xi\, y^\alpha\left(\frac{\rho_1}{\rho_1+\rho_3}\partial_{3\alpha}-\frac{\rho_3}{\rho_1+\rho_3}\partial_{2\alpha}\right)\\
+i\frac{(1-\xi)\rho_1}{\rho_1+\rho_2}\partial_{\go \alpha}\partial_1 {}^\alpha-i\left(\frac{(1-\xi)\rho_2}{\rho_1+\rho_2}+\frac{\xi\rho_3}{\rho_1+\rho_3}\right) \partial_{\go \alpha}\partial_2 {}^\alpha+i\frac{\xi\rho_1}{\rho_1+\rho_3}\partial_{\go \alpha}\partial_3 {}^\alpha\Big\} \go CCC,
\end{multline}
\begin{multline}
 \widehat{B}_3^{\eta\eta}\ast \go \approx-\frac{\eta^2}{4} \int_0^1 d\mathcal{T}\, \mathcal{T}  \int d^3 \rho_+ \delta\left(1-\sum_{i=1}^3 \rho_i\right)   \int_0^1 d\xi\, \frac{\rho_1\, \left[z_\alpha\left(y^\alpha+i\partial_{\go} {}^\alpha\right)\right]^2 e^{i\mathcal{T}\, z_\alpha y^\alpha} }{(\rho_1+\rho_2)(\rho_1+\rho_3)}\times\\
\times\exp\Big\{\mathcal{T} z^\alpha\Big(\partial_{\go \alpha}-(\rho_1+\rho_3)\partial_{1\alpha}+(\rho_2-\rho_3)\partial_{2\alpha}+(\rho_1+\rho_2)\partial_{3\alpha}\Big)+y^\alpha\partial_{\go \alpha}\\
+(1-\xi)y^\alpha\left(\frac{\rho_1}{\rho_1+\rho_2}\partial_{1\alpha}-\frac{\rho_2}{\rho_1+\rho_2}\partial_{2\alpha}\right)+\xi\, y^\alpha\left(\frac{\rho_1}{\rho_1+\rho_3}\partial_{3\alpha}-\frac{\rho_3}{\rho_1+\rho_3}\partial_{2\alpha}\right)\\
+i\frac{(1-\xi)\rho_1}{\rho_1+\rho_2}\partial_{1 \alpha}\partial_\go {}^\alpha-i\left(\frac{(1-\xi)\rho_2}{\rho_1+\rho_2}+\frac{\xi\rho_3}{\rho_1+\rho_3}\right) \partial_{2 \alpha}\partial_\go {}^\alpha+i\frac{\xi\rho_1}{\rho_1+\rho_3}\partial_{3 \alpha}\partial_\go {}^\alpha\Big\} CCC\go.
\end{multline}

\subsubsection{$B_2$ driven terms}

The terms resulting from $\dr_x$ differentiation of $B_2$ and its multiplication with $W_1$
 by virtue of \eqref{W1goCeta}, \eqref{W1Cgoeta}, \eqref{C2EOM} and \eqref{starPr} give
\begin{multline}
\dr_x B^{\eta\, loc}_2\big|_{\go CCC}\approx-\frac{i\eta^2}{4}\int_0^1 d\mathcal{T}
\int_0^1 d\xi\int d^3\rho_+\, \delta\left(1-\sum_{i=1}^3\rho_i\right)\left(z_\alpha y^\alpha\right)
\Big[\left(\mathcal{T}z^\alpha-\xi y^\alpha\right)\partial_{\go \alpha}\Big]\times\\
\times \exp\Big\{i\mathcal{T}z_\alpha y^\alpha+i(1-\rho_2)\partial_{\go \alpha}\partial_1 {}^\alpha
-i\rho_2 \partial_{\go \alpha}\partial_2 {}^\alpha+\mathcal{T}z^\alpha
\Big(-(\rho_1+\rho_2)\partial_{\go \alpha}
-\rho_1\partial_{1 \alpha}+(\rho_2+\rho_3)\partial_{2 \alpha}+\partial_{3\alpha}\Big)\\
+y^\alpha\Big(\xi(\rho_1+\rho_2)\partial_{\go \alpha}
+\xi\rho_1\partial_{1 \alpha}-\xi(\rho_2+\rho_3)\partial_{2 \alpha}+(1-\xi)\partial_{3\alpha}\Big) \Big\}\go CCC\,,
\end{multline}
\begin{multline}
\dr_x B^{\eta\, loc}_2 \big|_{C\go CC}\approx
-\frac{i\eta^2}{4}\int_0^1 d\mathcal{T}\int_0^1 d\xi\int d^3\rho_+\, \delta\left(1-\sum_{i=1}^3\rho_i\right)
\left(z_\alpha y^\alpha\right)\Big[(\mathcal{T}z^\alpha-\xi y^\alpha)\partial_{\go \alpha}\Big]\times\\
\times \exp\Big\{i\mathcal{T} z_\alpha y^\alpha+i\rho_2\partial_{1 \alpha}\partial_{\go} {}^\alpha
+i(1-\rho_2)\partial_{\go \alpha}\partial_{2} {}^\alpha
+\mathcal{T}z^\alpha\Big(-\rho_3 \partial_{1 \alpha}+\rho_1 \partial_{\go \alpha}
+(\rho_1+\rho_2)\partial_{2 \alpha}+\partial_{3\alpha}\Big)\\
+y^\alpha\Big(\xi \rho_3 \partial_{1 \alpha}-\xi\rho_1 \partial_{\go \alpha}
-\xi(\rho_1+\rho_2)\partial_{2 \alpha}+(1-\xi)\partial_{3\alpha}\Big)\Big\}C \go CC\\
-\frac{i\eta^2}{4}\int_0^1 d\mathcal{T}\int_0^1 d\xi
\int d^3\rho_+\, \delta\left(1-\sum_{i=1}^3 \rho_i\right)\left(z_\alpha y^\alpha\right)
\Big[(\mathcal{T}z^\alpha-\xi y^\alpha)\partial_{\go \alpha}\Big]\times\\
\times \exp\Big\{ i\mathcal{T}z_\alpha y^\alpha+i(1-\rho_2)\partial_{1 \alpha}\partial_{\go} {}^\alpha
+i\rho_2 \partial_{\go \alpha}\partial_{\go} {}^\alpha
+\mathcal{T}z^\alpha\Big(-(\rho_1+\rho_2)\partial_{1 \alpha}-\rho_1\partial_{\go\alpha}
+\rho_3\partial_{2 \alpha}+\partial_{3\alpha}\Big)\\
+y^\alpha\Big(\xi(\rho_1+\rho_2)\partial_{1 \alpha}
+\xi\rho_1 \partial_{\go \alpha}-\xi\rho_3 \partial_{2 \alpha}+(1-\xi)\partial_{3\alpha}\Big)\Big\}C\go CC\\
-\frac{i\eta^2}{4}\int_0^1 d\mathcal{T}\int_0^1 d\xi\int d^3 \rho_+\, \delta\left(1-\sum_{i=1}^3 \rho_i\right)
\left(z_\alpha y^\alpha\right)\Big[  \left(\mathcal{T}z^\alpha+ (1-\xi)y^\alpha\right) 
\partial_{\go \alpha}\Big]\times\\
\times \exp\Big\{i\mathcal{T} z_\alpha y^\alpha   +i(1- \rho_2) 
\partial_{\go \alpha}\partial_{2} {}^\alpha
-i\rho_2 \partial_{\go \alpha} \partial_3 {}^\alpha
+\mathcal{T}z^\alpha\Big(-\partial_{1 \alpha}-(\rho_1+\rho_2)\partial_{\go \alpha}
-\rho_1 \partial_{2 \alpha}+(\rho_2+\rho_3)\partial_{3\alpha}\Big)\\
+y^\alpha\Big(\xi\partial_{1\alpha}-(1-\xi)(\rho_1+\rho_2)\partial_{\go \alpha}
-(1-\xi)\rho_1 \partial_{2 \alpha}+(1-\xi)(\rho_2+\rho_3)\partial_{3\alpha}\Big)\Big\}C\go CC\,,
\end{multline}%
\begin{multline}
\dr_x B^{\eta\, loc}_2\big|_{CC\go C}\approx-\frac{i\eta^2}{4}\int_0^1 d\mathcal{T}
\int_0^1 d\xi \int d^3\rho_+ \, \delta\left(1-\sum_{i=1}^3\rho_i\right)\left(z_\alpha y^\alpha\right)
\Big[\left(\mathcal{T}z^\alpha-\xi y^\alpha\right)\partial_{\go \alpha}\Big]\times\\
\times \exp\Big\{i\mathcal{T} z_\alpha y^\alpha+i(1-\rho_2)\partial_{2\alpha}\partial_{\go} {}^\alpha
-i\rho_2 \partial_{1 \alpha}\partial_{\go} {}^\alpha+\mathcal{T}z^\alpha
\Big(-(\rho_1+\rho_2)\partial_{1 \alpha}+\rho_3 \partial_{2 \alpha}+(\rho_2+\rho_3)\partial_{\go \alpha}
+\partial_{3\alpha}\Big)\\
+y^\alpha\Big(\xi(\rho_1+\rho_2)\partial_{1 \alpha}-\xi\rho_3 \partial_{2 \alpha}
-\xi(\rho_2+\rho_3)\partial_{\go \alpha}+(1-\xi)\partial_{3\alpha}\Big)\Big\}CC\go C\\
-\frac{i\eta^2}{4}\int_0^1 d\mathcal{T} \int_0^1 d\xi \int d^3\rho_+ \,
 \delta\left(1-\sum_{i=1}^{3}\rho_i\right)\left(z_\alpha y^\alpha\right)
 \Big[\left(\mathcal{T}z^\alpha +(1-\xi)y^\alpha\right)\partial_{\go \alpha}\Big]\times\\
\times \exp\Big\{i\mathcal{T} z_\alpha y^\alpha+i\rho_2\partial_{2\alpha}\partial_{\go} {}^\alpha
+i(1-\rho_2)\partial_{\go \alpha}\partial_3 {}^\alpha+\mathcal{T}z^\alpha
\Big(-\partial_{1 \alpha}-\rho_3 \partial_{2 \alpha}+\rho_1\partial_{\go \alpha}
+(\rho_1+\rho_2)\partial_{3\alpha}\Big)\\
+y^\alpha \Big(\xi \partial_{1 \alpha}-(1-\xi)\rho_3\partial_{2 \alpha}
+(1-\xi)\rho_1\partial_{\go \alpha}+(1-\xi)(\rho_1+\rho_2)\partial_{3\alpha}\Big)\Big\}CC\go C\\
-\frac{i\eta^2}{4}\int_0^1 d\mathcal{T}\int_0^1 d\xi\int d^3 \rho_+\, \delta\left(1-\sum_{i=1}^3 \rho_i\right)
\left(z_\alpha y^\alpha\right)\Big[\left(\mathcal{T}z^\alpha+(1-\xi)y^\alpha\right)\partial_{\go \alpha}\Big]\times\\
\times \exp\Big\{i\mathcal{T} z_\alpha y^\alpha+i(1-\rho_2)\partial_{2 \alpha}\partial_{\go} {}^\alpha
+i\rho_2 \partial_{\go \alpha}\partial_3 {}^\alpha+\mathcal{T}z^\alpha
\Big(-\partial_{1 \alpha}-(\rho_1+\rho_2)\partial_{2 \alpha}
-\rho_1\partial_{\go \alpha}+\rho_3 \partial_{3\alpha}\Big)\\
+y^\alpha\Big(\xi\partial_{1 \alpha}-(1-\xi)(\rho_1+\rho_2)\partial_{2 \alpha}
-(1-\xi)\rho_1\partial_{\go \alpha}+(1-\xi)\rho_3\partial_{3\alpha}\Big)\Big\}CC\go C\,,
\end{multline}
 \begin{multline}
\dr_x B_2^{\eta\, loc}\big|_{CCC\go}\approx-\frac{i\eta^2}{4}\int_0^1 d\mathcal{T}\int_0^1 d\xi
\int d^3 \rho_+\, \delta\left(1-\sum_{i=1}^3\rho_i\right)\left(z_\alpha y^\alpha\right)
\Big[\left(\mathcal{T}z^\alpha+(1-\xi)y^\alpha\right)\partial_{\go \alpha}\Big]\times\\
\times \exp\Big\{i\mathcal{T} z_\alpha y^\alpha +i(1-\rho_2)\partial_{3\alpha}\partial_{\go} {}^\alpha
-i\rho_2 \partial_{2 \alpha}\partial_{\go} {}^\alpha
+\mathcal{T}z^\alpha\Big(-\partial_{1 \alpha}-(\rho_2+\rho_3)\partial_{2 \alpha}
+\rho_1 \partial_{3\alpha}+(\rho_1+\rho_2)\partial_{\go \alpha}\Big)\\
+y^\alpha\Big(\xi\partial_{1 \alpha}-(1-\xi)(\rho_2+\rho_3)\partial_{2 \alpha}
+(1-\xi)\rho_1 \partial_{3\alpha}+(1-\xi)(\rho_1+\rho_2)\partial_{\go \alpha}\Big)\Big\}CCC\go\,,
\end{multline}
\begin{multline}
W_{1\, \go C}^\eta \ast B_2^{\eta\, loc}\approx\frac{i\eta^2}{4}\int_0^1 d\mathcal{T}\T \int_0^1 d\sigma
\int d^3\rho_+\, \frac{\delta\left(1-\sum_{i=1}^3 \rho_i\right)}{\rho_1+\rho_2}\left(z^\gamma \partial_{\go \gamma}\right)\Big[z_\alpha y^\alpha+i\sigma z^\alpha \partial_{\go \alpha}\Big]\times\\
\times \exp\Big\{i\mathcal{T} z_\alpha y^\alpha+i(1-\sigma)\partial_{\go \alpha}\partial_1 {}^\alpha
-i\frac{\rho_1\sigma}{\rho_1+\rho_2} \partial_{\go \alpha}\partial_{2}{}^\alpha+i\frac{\rho_2\sigma}{\rho_1+\rho_2} \partial_{\go\alpha}\partial_3 {}^\alpha\\
+\mathcal{T}z^\alpha\Big(-(\rho_1+\rho_2+\sigma \rho_3)\partial_{\go \alpha}-(\rho_1+\rho_2)\partial_{1 \alpha}+(\rho_3-\rho_1)\partial_{2 \alpha}+(\rho_3+\rho_2)\partial_{3\alpha}\Big)\\
+y^\alpha\Big(\sigma\partial_{\go \alpha}-\frac{\rho_1}{\rho_1+\rho_2}\partial_{2 \alpha}+\frac{\rho_2}{\rho_1+\rho_2}\partial_{3\alpha}\Big)\Big\}\go CCC\,,
\end{multline}
 \begin{multline}
W_{1\, C\go}^\eta \ast B_2^{\eta \, loc}\approx\frac{i\eta^2}{4}\int_0^1 d\mathcal{T}\T
\int_0^1 d\sigma \int d^3\rho_+\, \frac{\delta\left(1-\sum_{i=1}^3 \rho_i\right)}{\rho_1+\rho_2}
\left(z^\gamma \partial_{\go \gamma}\right)\Big[z_\alpha y^\alpha+i\sigma z^\alpha \partial_{\go \alpha}\Big]\times\\
\times \exp\Big\{i\mathcal{T} z_\alpha y^\alpha
+i(1-\sigma)\partial_{1 \alpha}\partial_\go {}^\alpha
-i\frac{\rho_1\sigma}{\rho_1+\rho_2} \partial_{\go \alpha}\partial_{2}{}^\alpha
+i\frac{\rho_2\sigma}{\rho_1+\rho_2} \partial_{\go\alpha}\partial_3 {}^\alpha\\
+\mathcal{T}z^\alpha\Big(-(\rho_1+\rho_2+\sigma \rho_3)\partial_{\go \alpha}
-(\rho_1+\rho_2)\partial_{1 \alpha}+(\rho_3-\rho_1)\partial_{2 \alpha}+(\rho_3+\rho_2)\partial_{3\alpha}\Big)\\
+y^\alpha\Big(\sigma\partial_{\go \alpha}-\frac{\rho_1}{\rho_1+\rho_2}\partial_{2 \alpha}
+\frac{\rho_2}{\rho_1+\rho_2}\partial_{3\alpha}\Big)\Big\}C \go CC\,,
\end{multline}
 \begin{multline}
B_2^{\eta\, loc}\ast W_{1\, \go C}^\eta\approx\frac{i\eta^2}{4}\int_0^1 d\mathcal{T}\T
 \int_0^1 d\sigma \int d^3\rho_+\, \frac{\delta\left(1-\sum_{i=1}^3 \rho_i\right)}{\rho_1+\rho_2}
 \Big[z_\alpha y^\ga-i\sigma z^\alpha \partial_{\go\alpha}\Big]\left(z^\gamma \partial_{\go \gamma}\right)\times\\
\times \exp\Big\{i\mathcal{T} z_\alpha y^\alpha+i(1-\sigma)\partial_{\go \alpha}\partial_3 {}^\alpha
-i\frac{\rho_1 \sigma}{\rho_1+\rho_2}\partial_{1 \alpha}\partial_{\go} {}^\alpha
+i\frac{\rho_2 \sigma}{\rho_1+\rho_2}\partial_{2 \alpha}\partial_{\go} {}^\alpha\\
+\mathcal{T}z^\alpha\Big(-(\rho_3+\rho_1)\partial_{1 \alpha}-(\rho_3-\rho_2)\partial_{2 \alpha}
+(\rho_1+\rho_2-\sigma\rho_3)\partial_{\go \alpha}+(\rho_1+\rho_2)\partial_{3\alpha}\Big)\\
+y^\alpha\Big(\frac{\rho_1}{\rho_1+\rho_2}\partial_{1 \alpha}-\frac{\rho_2}{\rho_1+\rho_2}\partial_{2\alpha}
-\sigma\partial_{\go\alpha}\Big)\Big\}CC\go C\,,
\end{multline}
\begin{multline}
B_2^{\eta\, loc}\ast W_{1\, C\go}^\eta\approx\frac{i\eta^2}{4}\int_0^1 d\mathcal{T}\T
\int_0^1 d\sigma \int d^3\rho_+\, \frac{\delta\left(1-\sum_{i=1}^3 \rho_i\right)}{\rho_1+\rho_2}
\Big[z_\alpha y^\ga-i\sigma z^\alpha \partial_{\go\alpha}\Big]\left(z^\gamma \partial_{\go \gamma}\right)\times\\
\times \exp\Big\{i\mathcal{T} z_\alpha y^\alpha+i(1-\sigma)\partial_{3 \alpha}\partial_\go {}^\alpha
-i\frac{\rho_1 \sigma}{\rho_1+\rho_2}\partial_{1 \alpha}\partial_{\go} {}^\alpha
+i\frac{\rho_2 \sigma}{\rho_1+\rho_2}\partial_{2 \alpha}\partial_{\go} {}^\alpha\\
+\mathcal{T}z^\alpha\Big(-(\rho_3+\rho_1)\partial_{1 \alpha}-(\rho_3-\rho_2)\partial_{2 \alpha}+(\rho_1+\rho_2
-\sigma\rho_3)\partial_{\go \alpha}+(\rho_1+\rho_2)\partial_{3\alpha}\Big)\\
+y^\alpha\Big(\frac{\rho_1}{\rho_1+\rho_2}\partial_{1 \alpha}-\frac{\rho_2}{\rho_1+\rho_2}\partial_{2\alpha}
-\sigma\partial_{\go\alpha}\Big)\Big\}CCC\go.
\end{multline}

\subsubsection{$W_2$ driven terms}
Terms resulting from star product with $W_2^{\eta\eta}$  are
\begin{multline}
C\ast\widehat{W}_{2\, \go CC}^{\eta\eta}\approx\frac{\eta^2}{4}
\int_0^1 d\mathcal{T}\,\T \int d^4\rho_+\, \delta\left(1-\sum_{i=1}^4 \rho_i\right)
\frac{\rho_1\left(z^\gamma \partial_{\go \gamma}\right)^2}{(\rho_1+\rho_2)(\rho_3+\rho_4)}\times\\
\times \exp\Big\{i\mathcal{T}z_\alpha y^\alpha+\mathcal{T}z^\alpha
\Big(-\partial_{1 \alpha}+(1-\rho_2)\partial_{\go\alpha}-(\rho_3+\rho_4)\partial_{2\alpha}
+(\rho_1+\rho_2)\partial_{3 \alpha}\Big)+y^\alpha\partial_{1 \alpha}\\
+\frac{\rho_1\rho_3}{(\rho_1+\rho_2)(\rho_3+\rho_4)}
\left(y^\alpha \partial_{\go \alpha}+i\partial_{1 \alpha}\partial_{\go}{}^\alpha\right)
+i\left(\frac{(1-\rho_4)\rho_2}{\rho_1+\rho_2}+\rho_4\right)\partial_{\go \alpha}\partial_1 {}^\alpha
-i\frac{\rho_4\rho_1}{\rho_3+\rho_4}\partial_{\go \alpha}\partial_2 {}^\alpha\Big\}C \go CC,
\end{multline}
\begin{multline}
\widehat{W}_{2\, \go CC}^{\eta\eta}\ast C\approx\frac{\eta^2}{4}
\int_0^1 d\mathcal{T}\,\T \int d^4\rho_+\, \delta\left(1-\sum_{i=1}^4 \rho_i\right)
\frac{\rho_1 \left(z^\gamma \partial_{\go \gamma}\right)^2}{(\rho_1+\rho_2)(\rho_3+\rho_4)}\times\\
\times \exp\Big\{i\mathcal{T}z_\alpha y^\alpha+\mathcal{T}z^\alpha\Big((1-\rho_2)\partial_{\go\alpha}
-(\rho_3+\rho_4)\partial_{1\alpha}+(\rho_1+\rho_2)\partial_{2 \alpha}+\partial_{3 \alpha}\Big)
+y^\alpha\partial_{3 \alpha}\\
+\frac{\rho_1\rho_3}{(\rho_1+\rho_2)(\rho_3+\rho_4)}
\left(y^\alpha \partial_{\go \alpha}+i\partial_{\go \alpha}\partial_{3}{}^\alpha\right)
+i\left(\frac{(1-\rho_4)\rho_2}{\rho_1+\rho_2}+\rho_4\right)\partial_{\go \alpha}\partial_1 {}^\alpha
-i\frac{\rho_4\rho_1}{\rho_3+\rho_4}\partial_{\go \alpha}\partial_2 {}^\alpha\Big\} \go CC C,
\end{multline}

\begin{multline}
C\ast \widehat{W}_{2\, CC\go}^{\eta\eta}\approx\frac{\eta^2}{4}\int_0^1 d\mathcal{T}\, \mathcal{T}
\int d^4 \rho_+\, \delta\left(1-\sum_{i=1}^4 \rho_i\right)
\frac{\rho_1\left(z^\gamma \partial_{\go \gamma}\right)^2}{(\rho_1+\rho_2)(\rho_3+\rho_4)}\times\\
\times\exp\Big\{i\mathcal{T}z_\alpha y^\alpha+\mathcal{T}z^\alpha\Big(-\partial_{1 \alpha}
-(\rho_1+\rho_2)\partial_{2 \alpha}+(\rho_3+\rho_4)\partial_{3 \alpha}+(1-\rho_2)\partial_{\go \alpha}\Big)
+y^\alpha\partial_{1 \alpha}\\
+\frac{\rho_1\rho_3}{(\rho_1+\rho_2)(\rho_3+\rho_4)}\left(y^\alpha \partial_{\go \alpha}
+i\partial_{1 \alpha}\partial_{\go} {}^\alpha\right)
+i\left(\frac{(1-\rho_4)\rho_2}{\rho_1+\rho_2}+\rho_4\right)\partial_{2 \alpha}\partial_{\go} {}^\alpha
-i\frac{\rho_1\rho_4}{\rho_3+\rho_4}\partial_{1 \alpha}\partial_{\go} {}^\alpha\Big\}CCC\go,
\end{multline}
\begin{multline}
\widehat{W}_{2\, CC\go}^{\eta\eta}\ast C\approx\frac{\eta^2}{4}\int_0^1 d\mathcal{T}\, \mathcal{T}\int d^4 \rho_+\, \delta\left(1-\sum_{i=1}^4 \rho_i\right)\frac{\rho_1\left(z^\gamma \partial_{\go \gamma}\right)^2}{(\rho_1+\rho_2)(\rho_3+\rho_4)}\times\\
\times\exp\Big\{i\mathcal{T}z_\alpha y^\alpha+\mathcal{T}z^\alpha\Big(-(\rho_1+\rho_2)\partial_{1 \alpha}+(\rho_3+\rho_4)\partial_{2 \alpha}+(1-\rho_2)\partial_{\go \alpha}+\partial_{3 \alpha}\Big)+y^\alpha\partial_{3 \alpha}\\
+\frac{\rho_1\rho_3}{(\rho_1+\rho_2)(\rho_3+\rho_4)}\left(y^\alpha \partial_{\go \alpha}+i\partial_{\go \alpha}\partial_{3} {}^\alpha\right)+i\left(\frac{(1-\rho_4)\rho_2}{\rho_1+\rho_2}+\rho_4\right)\partial_{2 \alpha}\partial_{\go} {}^\alpha-i\frac{\rho_1\rho_4}{\rho_3+\rho_4}\partial_{1 \alpha}\partial_{\go} {}^\alpha\Big\}CC\go C,
\end{multline}
 \begin{multline}
C\ast \widehat{W}_{2\, C\go C}^{\eta \eta}\approx-\frac{\eta^2}{2}\int_0^1 d\mathcal{T}\, \mathcal{T}
\int d^4\rho_+\, \delta\left(1-\sum_{i=1}^4 \rho_i\right)
\frac{(\rho_1+\rho_3)\left(z^\gamma\partial_{\go \gamma}\right)^2 }{(\rho_1+\rho_2)(\rho_3+\rho_4)}\times\\
\times \exp\Big\{i\mathcal{T}z_\alpha y^\alpha
+\mathcal{T}z^\alpha\Big(-\partial_{1 \alpha}-(\rho_3+\rho_4)\partial_{2 \alpha}
+(\rho_1-\rho_3)\partial_{\go \alpha}+(\rho_1+\rho_2)\partial_{3 \alpha}\Big)
+y^\alpha \partial_{1 \alpha}\\-\frac{\rho_3\rho_1}{(\rho_1+\rho_2)(\rho_3+\rho_4)}
\left(y^\alpha \partial_{\go\alpha}+i\partial_{1 \alpha}\partial_{\go} {}^\alpha\right)
+i\frac{\rho_4(1-\rho_2)}{\rho_3+\rho_4}\partial_{\go \alpha}\partial_{2} {}^\alpha
+i\frac{\rho_2(1-\rho_4)}{\rho_1+\rho_2}\partial_{1 \alpha}\partial_\go {}^\alpha\Big\}CC\go C,
\end{multline}
\begin{multline}
\widehat{W}_{2\, C\go C}^{\eta \eta}\ast C\approx-\frac{\eta^2}{2}\int_0^1 d\mathcal{T}\, \mathcal{T}
 \int d^4\rho_+\, \delta\left(1-\sum_{i=1}^4 \rho_i\right)
 \frac{(\rho_1+\rho_3)\left(z^\gamma\partial_{\go \gamma}\right)^2}{(\rho_1+\rho_2)(\rho_3+\rho_4)}\times\\
\times \exp\Big\{i\mathcal{T}z_\alpha y^\alpha
+\mathcal{T}z^\alpha\Big(-(\rho_3+\rho_4)\partial_{1 \alpha}+(\rho_1-\rho_3)\partial_{\go \alpha}
+(\rho_1+\rho_2)\partial_{2 \alpha}+\partial_{3 \alpha}\Big)+y^\alpha \partial_{3 \alpha}
\\-\frac{\rho_3\rho_1}{(\rho_1+\rho_2)(\rho_3+\rho_4)}\left(y^\alpha \partial_{\go\alpha}
+i\partial_{\go \alpha}\partial_3 {}^\alpha\right)
+i\frac{\rho_4(1-\rho_2)}{\rho_3+\rho_4}\partial_{\go \alpha}\partial_{2} {}^\alpha
+i\frac{\rho_2(1-\rho_4)}{\rho_1+\rho_2}\partial_{1 \alpha}\partial_\go {}^\alpha\Big\}C\go CC.
\end{multline}

In the end of this section let us stress again that all terms on
the \rhs of vertex (\ref{rhs}) are free from $C$-field
contractions $\p_{i\ga}\p_j^\ga$  in the exponentials, hence being
spin-local. This is the central result of this paper.

\section{Conclusion}\label{Conclussion}

In this paper we have analyzed the  $\go C^3$ vertices in the
equation for the zero-form $C$ (\ref{HSsketch2}) in the
holomorphic $\eta^2$ sector, showing that these vertices are
spin-local in the terminology of \cite{2a2}. In particular, they
contain the holomorphic part of the $\phi^4$ vertex in the
Lagrangian nomenclature for a spin-zero scalar field $\phi$. This
is another step in the analysis of locality of HS gauge theory
performed in \cite{4a1,4a2}. To complete the analysis of
spin-locality of the HS gauge theory at quartic order it remains
to extend these results to the mixed $\eta\bar \eta$ sector. This
problem differs in some respects from the (anti)holomorphic one
and will be analyzed elsewhere.

On the other hand, there are remaining problems even in the
holomorphic sector left unsolved. The most important one is to
find explicit $Z$-independent local form of the holomorphic vertex
$\go C^3$. The naive attempt to set $Z=0$ in the vertex obtained
in this paper does not necessarily lead to correct result since
the omitted terms in $\Hp_0$ are needed for consistency of the
equations and may contribute to the sector of equations. Indeed,
setting $Z=0$ corresponds to the application of the conventional
homotopy projector which does not eliminate the part of the vertex
in $\Hp$. Let us stress again in this regard that the elaborated
technique based on dropping off terms from $\Hp_0$ turns out to be
highly efficient for checking out spin-locality. To obtain
explicit form of these vertices there are two alternative ways of
the analysis.

One is to eliminate the $Z$-dependence from the vertex by direct partial integration.
  Being technically involved and not at all obvious  due to the
need of using Schouten identity and partial integrations, this program is realised
at least for a particular vertex in the forthcoming paper \cite{GelKor}.

Another is to apply the limiting shifted homotopy procedure with
appropriately chosen shift when solving for HS fields. Note that
the $Z$-dependence for HS master fields has been found using no
shifted homotopies in our paper. It would be interesting to
understand if the locality is reached within well elaborated contracting
homotopy approach. Since the choice of homotopy shift and hence cohomology
projector via resolution of identity (\ref{1}), (\ref{projector}) affects field redefinitions that can
themselves be non-local the art is to find a shift that makes the
result manifestly spin-local. This is an interesting problem for the
future.

To summarize, the results of this paper indicate that equations of motion of
HS gauge theory have a tendency  of being spin-local. At this stage it is
crucially important to see whether this property extends to the mixed
$\eta\bar\eta \go C^3$ sector of equation (\ref{HSsketch2}) which is the most
urgent problem on the agenda.

\newcounter{appendix}
\setcounter{appendix}{1}
\renewcommand{\theequation}{\Alph{appendix}.\arabic{equation}}
\addtocounter{section}{1} \setcounter{equation}{0}
 \renewcommand{\thesection}{\Alph{appendix}.}
 \addcontentsline{toc}{section}{\,\,\,\,\,\,\,Appendix A.  Useful  formulas}

\section*{Acknowledgments}
This work was supported by
the Russian Science Foundation grant 18-12-00507.

\section*{Appendix A. Useful  formulas}\label{Appendix}
Useful multiplication formula for the star product of functions of the form
\begin{equation}
f_j(z,y)=\int_0^1 d\tau_j \exp {i(\tau_j\, z_\alpha y^\alpha)} \phi_j(\tau_j z,(1-\tau_j)y\vert \tau_j \theta,\tau_j)
\end{equation}
is \cite{Vasiliev:2015wma}
\begin{multline}\label{starPr}
f_1\ast f_2 (z,y)=\int_0^1 d\tau_1 \int_0^1 d\tau_2 \int e^{iu_\alpha v^\alpha}\,
\exp {i( \tau_1 \circ \tau_2 z_\alpha y^\alpha)}\times\\
\phi_1\Big(\tau_1\big[(1-\tau_2)z-\tau_2 y+u\big],(1-\tau_1)\big[(1-\tau_2)y-\tau_2 z+u\big]\big| \tau_1 \theta,\tau_1\Big)\times\\
\phi_2\Big(\tau_2\big[(1-\tau_1)z+\tau_1 y-v\big]\big],(1-\tau_2)\big[(1-\tau_1)y+\tau_1 z+v\big]\big| \tau_2 \theta,\tau_2\Big)\,.
\end{multline}
For instance, if one function is  $z$-independent the following formulas are handy in star-product computation
\begin{equation}\label{YL}
f(y)\ast \Gamma(z,y)=f(y)\Gamma(z+i\overleftarrow{\partial}_f,y-i\overleftarrow{\partial}_f)\, ,
\end{equation}
\begin{equation}\label{YR}
\Gamma(z,y)\ast f(y)=\Gamma(z+i\partial_f,y+i\partial_f)f(y)\,.
\end{equation}

\bigskip

Second-order zero-form vertices are \cite{4a1}
\begin{multline}
\Upsilon^\eta_{\go CC}=-\frac{i\eta}{2}\int d^3\tau_+\, \delta\left(1-\sum_{i=1}^3 \tau_i\right)
\partial_{\omega\alpha}\left(\partial_1^\alpha+\partial_2^\alpha\right)
\exp\Big\{i(1-\tau_2)\partial_{\omega \alpha}\partial_1^\alpha
-i\tau_2 \partial_{\omega \alpha}\partial_2^\alpha\Big\}\\
\times \omega\big((1-\tau_3)y,\bar{y}\big)\ust C\big(\tau_1 y,\bar{y}\big)\ust C\big((\tau_1-1)y,\bar{y}\big)k,
\end{multline}
\begin{multline}
\Upsilon_{CC\go}^\eta=-\frac{i\eta}{2}\int d^3\tau_+ \,\delta\left(1-\sum_{i=1}^3 \tau_i\right)
\partial_{\omega\alpha}\left(\partial_1^\alpha+\partial_2^\alpha\right)
\exp\Big\{i(1-\tau_1)\partial_{2 \alpha}\partial_{\go}^\alpha
-i\tau_1\partial_{1\alpha} \partial_{\go}^\alpha\Big\}\\
\times C\big((1-\tau_2)y,\bar{y}\big)\ust C\big(-\tau_2 y,\bar{y}\big)\ust \omega\big((\tau_3-1)y,\bar{y}\big)k,
\end{multline}
\begin{multline}
\Upsilon^\eta_{C\go C}=-\frac{i\eta}{2}\int d^3 \tau_+\, \delta\left(1-\sum_{i=1}^3 \tau_i\right)
\partial_{\omega\alpha}\left(\partial_1^\alpha+\partial_2^\alpha\right)
\exp\Big\{i\tau_3 \partial_{1\alpha}\partial_{\go}^\alpha
+i(1-\tau_3)\partial_{\omega \alpha}\partial_2^\alpha\Big\}\\
\times C\big(\tau_2 y,\bar{y}\big)\ust \omega\big(-\tau_1 y,\bar{y}\big)\ust C\big((\tau_2-1)y,\bar{y}\big)k\\
-\frac{i\eta}{2}\int d^3 \tau_+\, \delta\left(1-\sum_{i=1}^3 \tau_i\right)\partial_{\omega\alpha}
\left(\partial_1^\alpha+\partial_2^\alpha\right)\exp\Big\{i(1-\tau_2)\partial_{1\alpha}\partial_{\go}^\alpha
+i\tau_2 \partial_{\omega \alpha}\partial_2^\alpha\Big\}\\
\times C\big((1-\tau_3)y,\bar{y}\big)\ust \omega\big(\tau_1 y,\bar{y}\big)\ust C\big(-\tau_3 y,\bar{y}\big)k\,.
\end{multline}

\addtocounter{appendix}{1}
\renewcommand{\theequation}{\Alph{appendix}.\arabic{equation}}
\addtocounter{section}{1} \setcounter{equation}{0}
 \addcontentsline{toc}{section}{\,\,\,\,\,\,\,Appendix B.  $B_3^{\eta\eta}$}

\section*{Appendix B.  $B_3^{\eta\eta}$}
\label{SecB3}
Computation of $B^{\eta\eta}_3$ goes as follows.
Equation for $B_3^{\eta\eta}$ from \eqref{HS5} is
\begin{equation}
2i\dr_z B_3^{\eta\eta}=[S_1^\eta,B_2]_\ast+[S_2^{\eta\eta},C]_\ast\,.
\end{equation}

An important observation of Section 6.2 of \cite{4a2} based on the
 technique of re-ordering operators $O_{\gb}f(z,y)$  was that
if $S_2^{\eta\eta}$
is computed using $B_2^{\eta\, loc}$  \eqref{B2loc}, then the
contribution to    the vertices $ \Upsilon^{\eta\eta}(\go,\go,C,C)$
from   $S^{\eta\eta}_2$  vanishes at $\gb\to-\infty$. Proceeding analogously one can see that
contribution to   the vertices $ \Upsilon^{\eta\eta}(\go,C,C,C)$ from such $S_2^{\eta\eta}$ also
  vanishes at $\gb\to-\infty$.

Hence to find the part of $B_3^{\eta\eta}$ that contributes to
$\widehat{\Upsilon}^{\eta\eta}$ one has to solve the equation
\begin{equation}
\dr_z \widehat{B}_3^{\eta\eta}=\frac{i}{2}[B_2^{\eta\, loc},S_1^\eta]_\ast.
\end{equation}

$S_1^\eta$ is given by \eqref{S1vvedenie}. Replacing the
integration over simplex in  \eqref{B2loc} by the integration over
unit square
\begin{equation}\label{simp_to_square1}
\int d^3\tau_+ \delta(1-\sum_{i=1}^3\tau_i)=\int_0^1 d\tau_1\int_0^1 d\sigma (1-\tau_1),
\end{equation}
by changing the variables as follows
\begin{equation}\label{simp_to_square2}
\tau_2=(1-\tau_1)\sigma,\;\; \tau_3=(1-\tau_1)(1-\sigma)
\end{equation} and then performing partial integration    with respect to $\gt_1$
 using star-product formula \eq{starPr} and
   dropping the terms from $\Hp_1$ we obtain
\begin{multline}\label{commut1}
\left[B_2^{\eta\, loc} ,S_1^\eta\right]_\ast\approx-\frac{i\eta^2}{2}\theta^\beta
\int_0^1 d\tau_1 \int_0^1 dt\int_0^1 d\sigma \, t(1-t)(1-\tau_1)e^{i\tau_1\circ t \,z_\alpha y^\alpha}
(z_\alpha y^\alpha) z_\beta\times\\
\Bigg\{C\Big(\big[-\tau_1(1-t)-\sigma t(1-\tau_1)\big]z+\sigma y\Big)C
\Big(\big[-\tau_1(1-t)+t(1-\tau_1)(1-\sigma)\big]z-(1-\sigma)y\Big)C\Big(t(1-\tau_1)z\Big)\\
-C\Big(-t(1-\tau_1)z\Big)C\Big(\big[\tau_1(1-t)-\sigma t(1-\tau_1)\big]z-\sigma y\Big)C
\Big(\big[\tau_1(1-t)+(1-\sigma)(1-\tau_1)t\big]z+(1-\sigma)y\Big)
\Bigg\}.
\end{multline}
Recall that we use notation with hidden $\bar y$ variables:
\bee\label{ustbybz}&& C(-\tau_1 z+\sigma(1-\tau_1) y) C(-\tau_1 z-(1-\sigma)(1-\tau_1)y)\equiv
\\ \nn &&\equiv
C(-\tau_1 z+\sigma(1-\tau_1) y,\by)\ust  C(-\tau_1 z-(1-\sigma)(1-\tau_1)y,\by)\,.
\eee

Since only small values of \be \label{circ}\mathcal{T}:=\tau_1\circ t=\tau_1(1-t)+t(1-\tau_1)\ee
contribute to $\widehat{\Upsilon}^{\eta\eta}$ one needs to consider two triangle regions of the init
square in $(\tau_1,t)$ coordinates.
 Only the lower triangle with small $t$ and $\tau_1$ contributes because the upper-one gives $\mathcal{T}^3$ in the
pre-exponential  thus belonging to $\Hp_1$. The following change of variables is handy
in the further analysis
\begin{multline}\label{lowTr1}
\int d\mathcal{T} \int d\tau_1\, dt\, \theta(\tau_1)\theta(t)\theta(\varepsilon-\tau_1-t) \delta(\mathcal{T}-\tau_1-t)f(\tau_1,t)=\\
=\int_0^\varepsilon d\mathcal{T} \int_0^\mathcal{T} dt\,
f(\mathcal{T}-t,t) =\int_0^\varepsilon d \mathcal{T} \int_0^1 dt^\prime \,
\mathcal{T}\, f(\mathcal{T}(1-t^\prime),\mathcal{T} t^\prime).
\end{multline}
Adding the terms from $\Hp$, which do not affect the HS field equations,
 one can reach further  simplifications.
For instance, one can add $\int_{\varepsilon}^{1}d\T\int_0^1 dt^\prime f(\T(1-t^\prime),\T t^\prime)$ to \eqref{lowTr1}, \ie
\begin{multline}\label{lowTr}
\int d\mathcal{T} \int d\tau_1\, dt\, \theta(\tau_1)\theta(t)\theta(\varepsilon-\tau_1-t) \delta(\mathcal{T}-\tau_1-t)f(\tau_1,t)\approx\\
\approx\int_0^1 d \mathcal{T} \int_0^1 dt^\prime \,
\mathcal{T}\, f(\mathcal{T}(1-t^\prime),\mathcal{T} t^\prime)\,.
\end{multline}
(Recall that sign $\approx$ means that equality is up to terms from $\Hp$.)

In \eqref{commut1} it is convenient to introduce new variables
\begin{equation}\label{simplex1}
\rho_1=t^\prime\sigma\, ,\;\;\; \rho_2=t^\prime(1-\sigma)\,,\;\;\; \rho_3=1-t^\prime.
\end{equation}
They form a simplex since $\rho_1+\rho_2+\rho_3=1$. The inverse  formulas are
\begin{equation}\label{simplex2}
\sigma=\frac{\rho_1}{\rho_1+\rho_2}\, ,\;\;\; (1-\sigma)=\frac{\rho_2}{\rho_1+\rho_2}\, ,\;\;\; t^\prime=1-\rho_3=\rho_1+\rho_2.
\end{equation}

In these new variables, the commutator takes the form
\begin{multline}\label{commutunif}
\left[ B_2^{\eta\, loc} ,S_1^\eta
\right]_\ast\approx-\frac{i\eta^2}{2}\theta^\beta z_\beta
(z_\alpha y^\alpha) \int_0^1 d\mathcal{T}\, \mathcal{T}^2 \int d^3
\rho_+ \delta\left(1-\sum_{i=1}^3 \rho_i\right)
e^{i\mathcal{T}\, z_\alpha y^\alpha}\times\\
\Big\{C\Big(-\mathcal{T}(\rho_1+\rho_3)z+\frac{\rho_1}{\rho_1+\rho_2}y\Big)C
\Big(\mathcal{T}(\rho_2-\rho_3)z-\frac{\rho_2}{\rho_1+\rho_2} y\Big)C\Big(\mathcal{T}(\rho_1+\rho_2)z\Big)\\
-C\Big(-\mathcal{T}(\rho_1+\rho_3)z\Big)C\Big(\mathcal{T}(\rho_2-\rho_3)z-\frac{\rho_3}{\rho_1+\rho_3}y\Big)
C\Big(\mathcal{T}(\rho_1+\rho_2)z+\frac{\rho_1}{\rho_1+\rho_3}y\Big)\Big\}.
\end{multline}

Note that $z$-dependence is the same in the both terms.
Introducing an additional integration parameter $\xi$, the 
$y$-dependence can be uniformized as follows
 \begin{multline}
\left[ B_2^{\eta\, loc} ,S_1^\eta
\right]_*\approx\frac{i\eta^2}{2}\theta^\beta z_\beta(z_\alpha y^\alpha)
\int_0^1 d\mathcal{T}\, \mathcal{T}^2\int d^3 \rho_+
\delta\left(1-\sum_{i=1}^3 \rho_i\right)
 \int_0^1 d\xi\,  \frac{\partial}{\partial \xi}
 \exp \big\{  \mathcal{Z}\big\}CCC\,,
\end{multline}
where the following notations are used
\begin{equation}
D_\alpha=-(\rho_1+\rho_3)\partial_{1\alpha}+(\rho_2-\rho_3)\partial_{2\alpha}+(\rho_1+\rho_2)\partial_{3\alpha},
\end{equation}
\begin{equation}
\mathcal{Z}=i\T z_\alpha y^\alpha + \mathcal{T} z^\alpha D_\ga +(1-\xi)y^\alpha\left(\frac{\rho_1}{\rho_1+\rho_2}\partial_{1\alpha}-\frac{\rho_2}{\rho_1+\rho_2}\partial_{2\alpha}\right)+\xi\, y^\alpha\left(\frac{\rho_1}{\rho_1+\rho_3}\partial_{3\alpha}-\frac{\rho_3}{\rho_1+\rho_3}
\partial_{2\alpha}\right)\,.
\end{equation}

Evaluating the derivative with respect to $\xi$ taking into account  that
\begin{equation}
\frac{\partial \mathcal{Z} }{\partial \xi}=\frac{\rho_1}{\T(\rho_1+\rho_2)(\rho_1+\rho_3)}y^\alpha \frac{\partial
\mathcal{Z}}{\partial  z^\alpha}
\end{equation}
along with the Schouten  identity
\begin{multline}
(\theta^\beta z_\beta)\left(y^\alpha \frac{\partial \mathcal{Z}}{\partial z^\alpha}\right)=
(\theta^\beta y_\beta)\left(z^\alpha \frac{\partial \mathcal{Z}}{\partial z^\alpha}\right)
+(z_\alpha y^\alpha)\left(\theta^\beta \frac{\partial \mathcal{Z}}{\partial z^\beta}\right)=\\
=(\theta^\beta y_\beta)\left(\T \frac{\partial \mathcal{Z}}{\partial \T}\right)
+(z_\alpha y^\alpha)\dr_z \mathcal{Z},
\end{multline}
the expression for the commutator can be rewritten in the  form
 \begin{multline}
\left[ B_2^{\eta\, loc} ,S_1^\eta
\right]_\ast\approx\mathrm{d}_z\left[ -\frac{\eta^2(z_\alpha
y^\alpha)^2}{2} \int_0^1 d\mathcal{T}\, \mathcal{T}\int d^3 \rho_+
\delta\left(1-\sum_{i=1}^3 \rho_i\right)
   \int_0^1 d\xi\, \frac{\rho_1\, \exp  \big\{ \mathcal{Z}\big\}}{(\rho_1+\rho_2)(\rho_1+\rho_3)}
      CCC\rule{0pt}{18pt}\right]\\
-\frac{\eta^2}{2} \theta^\beta y_\beta \int d\mathcal{T}\, \delta\left(1-\mathcal{T}\right)
\mathcal{T}^2 \int d^3 \rho_+ \delta\left(1-\sum_{i=1}^3 \rho_i\right)
\int_0^1 d\xi\, \frac{\rho_1\, (z_\alpha y^\alpha)  \exp  \big\{ \mathcal{Z}\big\} }{(\rho_1+\rho_2)(\rho_1+\rho_3)}CCC.
\end{multline}
Since the second (boundary) term belongs to $\Hp_1$ and thus contributes  to
 ${\Upsilon}^{\eta\eta}_+$   the part of $B_3^{\eta\eta}$ that
contributes to $\widehat{\Upsilon}^{\eta\eta}$ can be chosen in the form \eqref{B3final}.

\addtocounter{appendix}{1}
\renewcommand{\theequation}{\Alph{appendix}.\arabic{equation}}
\addtocounter{section}{1} \setcounter{equation}{0}
 \addcontentsline{toc}{section}{\,\,\,\,\,\,\,Appendix C. $W_2^{\eta\eta}$  }
 \renewcommand{\thesubsection}{\Alph{appendix}.\arabic{subsection}}

\section*{Appendix C. $W_2^{\eta\eta}$}
\label{W}

A particular solution for  $W^{\eta\eta}_2$ was found in
\cite{4a2}  where it was used in the computation of vertices
$\Upsilon^{\eta\eta}(\go,\go, C,C)$. However, this solution turns
out to be technically inconvenient for the analysis modulo $\Hp$
subspace. In this section we apply the approach proposed in the
previous section that allows us to single out the $\Hp_0$ part
from  $W_2^{\eta\eta}$. Vertices
$\Upsilon^{\eta\eta}(\go,\go,C,C)$ computed with $W_2^{\eta\eta}$
given by \eqref{W2CCgofinal}-\eqref{W2CgoCfinal} may differ from
those of  \cite{4a2}  by a local field redefinition.

To compute $W_2^{\eta\eta}$ consider the equation
\begin{multline}
\dr_x S_1^\eta+W_1^\eta\ast S_1^\eta+S_1^\eta\ast W_1^\eta+\dr_x S_2^{\eta\eta}+\omega\ast S_2^{\eta\eta}+S_2^{\eta\eta}\ast \omega+S_0\ast W_2^{\eta\eta}+W_2^{\eta\eta}\ast S_0=0.
\end{multline}
{As   mentioned in Appendix B
 contribution to   the vertices $ \Upsilon^{\eta\eta}(\go,C,C,C)$ from $S_2^{\eta\eta}$
  vanishes at $\gb\to-\infty$. The remaining equation to be solved  is
\begin{equation}
2i\dr_z \widehat{W}_2\approx\dr_x S_1^\eta+W_1^\eta\ast S_1^\eta+S_1^\eta\ast W_1^\eta.
\end{equation}
Here $S_1^\eta$ is given by \eqref{S1vvedenie} while $W_1^\eta$ consists of two parts \eqref{W1goCeta} and \eqref{W1Cgoeta}.

To calculate $\dr_x S_1^\eta$ one needs second-order zero-form vertices $\Upsilon^\eta(\go, C,C)$
obtained in \cite{4a1} with
additional shifts $\delta\Upsilon^\eta(\go, C,C)$ generated by the local shift $\delta B_2^\eta$
\eq{redef}. These are given by \eqref{VgoCC}-\eqref{VCgoC}.

\subsection{$W_{2\, CC\go}^{\eta\eta}$}
\label{SecW2CCgo}
Equation for $W_{2\, CC\go}^{\eta\eta}$ has the form
\begin{equation}\label{dzW2=}
2i\dr_z \widehat{W}_{2\, CC\go}^{\eta\eta}\approx\dr_x S_1^\eta\Big|_{CC\go}+S_1^\eta \ast W_{1\, C\go}^\eta.
\end{equation}
Computing $S_1^\eta\ast W^\eta_{1\, C\go}$ and dropping terms from $\Hp_1$ one obtains
discarding barred variables as in \eqref{ustbybz}
 \begin{multline}
S_1^\eta \ast W^\eta_{1\, C\go}\approx \frac{\eta^2}{2}\int_0^1 dt \int_0^1 d\tau_1 \int_0^1 d\sigma\,t(1-t)(1-\tau_1)^2 \left(\theta^\beta z_\beta\right)\left(z^\alpha \partial_{\omega \alpha}\right)\exp\Big\{i\tau_1 \circ t z_\alpha y^\alpha+i(1-\sigma)\partial_{2 \alpha}\partial_{\go}{}^\alpha\Big\}\times\\
\times C\Big(-t(1-\tau_1)z\Big)C\Big(\tau_1(1-t)z\Big)\omega\Big(\big[\tau_1(1-t)+
\sigma t(1-\tau_1)\big]z+\sigma y\Big)\,.
\end{multline}
Taking into account that only small values of $\tau_1 \circ t$ contribute to $\widehat{\Upsilon}^{\eta\eta}$ one can change integration variables as in \eqref{lowTr} to obtain
\begin{multline}
S_1^\eta \ast W_{1\, C\go}^\eta\approx\frac{\eta^2}{2}\int_0^1 d\mathcal{T}\, \mathcal{T}^2 \int_0^1 dt^\prime \int_0^1 d\sigma\, t^\prime\, \left(\theta^\beta z_\beta\right)\left(z^\alpha \partial_{\omega \alpha}\right)\exp\Big\{i\mathcal{T} z_\alpha y^\alpha+i(1-\sigma)\partial_{2 \alpha}\partial_{\go} {}^\alpha\Big\}\times\\
\times C\Big(-\mathcal{T} t^\prime z\Big)C\Big(\mathcal{T}(1-t^\prime)z\Big)\omega\Big(\mathcal{T}\big[(1-t^\prime)+t^\prime \sigma\big]z+\sigma y\Big)\,.
\end{multline}
In the simplex  variables \eqref{simplex1}, \eqref{simplex2} this expression
can be rewritten as
\begin{multline}\label{S1W1cgo}
S_1^\eta \ast W_{1\, C\go}^\eta\approx\frac{\eta^2}{2}\int_0^1 d\mathcal{T}\, \mathcal{T}^2 \int d^3 \rho_+\, \delta\left(1-\sum_{i=1}^3 \rho_i\right)\left(\theta^\beta z_\beta\right)\left(z^\alpha \partial_{\omega \alpha}\right)\exp\Big\{i\mathcal{T} z_\alpha y^\alpha+i\frac{\rho_2}{\rho_1+\rho_2}\partial_{2 \alpha}\partial_{\go} {}^\alpha\Big\}\times\\
\times C\Big(-\mathcal{T} (\rho_1+\rho_2) z\Big)C\Big(\mathcal{T}\rho_3 z\Big)\omega\Big(\mathcal{T}(\rho_1+\rho_3)z+\frac{\rho_1}{\rho_1+\rho_2} y\Big)\,.
\end{multline}
To compute $\dr_x S_1^\eta$ one has to use  vertex \eqref{VCCgo}
\begin{multline}\label{dxS1}
\dr_x S_1^\eta \big|_{CC\go}\approx\\
\approx-\frac{\eta^2}{2}\int_0^1 d\mathcal{T}\, \mathcal{T}^2 \int d^3 \rho_+\, \delta\left(1-\sum_{i=1}^3 \rho_i\right)\left(\theta^\beta z_\beta\right)\left(z^\alpha \partial_{\omega \alpha}\right)\exp\Big\{i\mathcal{T} z_\alpha y^\alpha+i(1-\rho_1)\partial_{2\alpha}\partial_{\go}{}^\alpha-i\rho_1 \partial_{1\alpha}\partial_{\go} {}^\alpha\Big\}\times\\
\times C\Big(-\mathcal{T}(\rho_1+\rho_2)z\Big)C\Big(\mathcal{T}\rho_3 z\Big)\go\Big(\mathcal{T}(\rho_1+\rho_3)z\Big).
\end{multline}

As in  Appendix B,  $y$-dependence in \eqref{dzW2=} can be uniformized with the help of the new integration parameter $\xi$.
Using new notation for brevity
\begin{multline}
\mathsf{Z}=i\mathcal{T}z_\alpha y^\alpha+\mathcal{T}z^\alpha\Big(-(\rho_1+\rho_2)\partial_{1\alpha}+\rho_3 \partial_{2 \alpha}+(\rho_1+\rho_3)\partial_{\omega \alpha}\Big)\\
+\xi\left(i\frac{\rho_2}{\rho_1+\rho_2}\partial_{2 \alpha}\partial_{\go}{}^\alpha+\frac{\rho_1}{\rho_1+\rho_2}y^\alpha \partial_{\omega \alpha}\right)+(1-\xi)\Big(i(1-\rho_1)\partial_{2 \alpha}\partial_{\go}{}^\alpha-i\rho_1 \partial_{1\alpha}\partial_{\go}{}^\alpha\Big)
\end{multline}
one has
from \eqref{dzW2=} taking into account \eqref{S1W1cgo} and \eqref{dxS1}
\begin{equation}
2i\dr_z \widehat{W}_{2\,
CC\go}^{\eta\eta}=\frac{\eta^2}{2}\int_0^1 d\mathcal{T}\,
\mathcal{T}^2 \int d^3 \rho_+\, \delta\left(1-\sum_{i=1}^3
\rho_i\right)\left(\theta^\beta z_\beta\right)\left(z^\alpha
\partial_{\omega \alpha}\right)\int_0^1 d\xi \,
\frac{\partial}{\partial \xi} e^{\mathsf{Z}} \, CC\go.
\end{equation}
Evaluating the derivative over $\xi$
\begin{equation}
\frac{\partial \mathsf{Z}}{\partial \xi} =\frac{\rho_1}{\rho_1+\rho_2}\Big(-i\rho_3 \partial_{2 \alpha}\partial_{\go}{}^\alpha+i(\rho_1+\rho_2)\partial_{1\alpha}\partial_\go {}^\alpha+y^\alpha\partial_{\omega \alpha}\Big)
\end{equation}
and taking into account  that
\bee\label{w2dz} \frac{\partial \mathsf{Z}}{\partial \xi}
 = \frac{-i\rho_1\, }{\T (\rho_1+\rho_2) } \partial_{\go}{}^\alpha  \frac{\partial \mathsf{Z}}{\partial z^\ga}
 \eee
along with the Schouten  identity
\begin{equation}\label{SchoutenW}
\left(\theta^\beta z_\beta\right)\left(\partial_\go{}^\alpha \frac{\partial \mathsf{Z}}{\partial z^\alpha}\right)=\left(\theta^\beta \partial_{\omega \beta}\right)\left(z^\alpha \frac{\partial \mathsf{Z}}{\partial z^\alpha}\right)-\left(\theta^\beta \frac{\partial \mathsf{Z}}{\partial z^\beta}\right)\left(z^\alpha \partial_{\go \alpha}\right)
\q\end{equation}
the pre-exponential part  can be written in the form
\begin{equation}
\left(\theta^\beta z_\beta\right)\left(\partial_\go{}^\alpha \frac{\partial \mathsf{Z}}{\partial z^\alpha}\right)=\left(\theta^\beta \partial_{\omega \beta}\right)\left(\T\frac{\partial \mathsf{Z}}{\partial \T} \right)-\left(z^\alpha \partial_{\go \alpha}\right)\dr_z \mathsf{Z}
\end{equation}
and thus \rhs of \eqref{dzW2=} can be put into the form
\begin{multline}
2i\dr_z \widehat{W}_{2\, CC\go}^{\eta\eta}=-\frac{i\eta^2}{2}\int_0^1 d\mathcal{T}\, \frac{\rho_1\mathcal{T}^2}{\rho_1+\rho_2} \int d^3\rho_+\, \delta\left(1-\sum_{i=1}^3 \rho_i\right)\int_0^1 d\xi\, \left(\theta^\beta \partial_{\omega \beta}\right)\left(z^\gamma \partial_{\go \gamma}\right)\frac{\partial}{\partial \mathcal{T}}e^{\mathsf{Z}}CC\go\\
+\frac{i\eta^2}{2}\int_0^1 d\mathcal{T} \, \frac{\rho_1
\mathcal{T}^2}{\rho_1+\rho_2}\int d^3\rho_+\,
\delta\left(1-\sum_{i=1}^3 \rho_i\right)\int_0^1 d\xi
\left(z^\gamma \partial_{\go \gamma}\right)^2\dr_z e^{\mathsf{Z}}
CC\go.
\end{multline}
After integrating by parts with respect to $\T$ in the first term, the resulting boundary term belongs to $\Hp$  (cf. the second case of \eqref{kernels}) and hence can be discarded. This brings equation for $\widehat{W}_{2\, CC\go}^{\eta\eta}$ to the form
\begin{equation}
2i\dr_z \widehat{W}_{2\, CC\go}^{\eta \eta} \approx\dr_z
\left\{\frac{i\eta^2}{2}\int_0^1 d\mathcal{T}\int_0^1 d\xi \int
d^3\rho_+ \, \delta\left(1-\sum_{i=1}^3
\rho_i\right)\frac{\mathcal{T}\rho_1 \left(z^\gamma \partial_{\go
\gamma}\right)^2}{\rho_1+\rho_2}e^{\mathsf{Z}}CC\go \right\}.
\end{equation}
This allows us to choose the part of $W_{2\, CC\go}^{\eta\eta}$,
that contributes to $\widehat{\Upsilon}^{\eta\eta}$, in the form
\begin{multline}
\widehat{W}_{2\, CC\go}^{\eta \eta}=\frac{\eta^2}{4}\int_0^1 d\mathcal{T} \int_0^1 d\xi\int d^3\rho_+\, \delta\left(1-\sum_{i=1}^3\rho_i\right)\frac{\mathcal{T}\rho_1 \left(z^\gamma \partial_{\go \gamma}\right)^2}{\rho_1+\rho_2}\times\\
\times\exp\Big\{i\mathcal{T}z_\alpha y^\alpha+\mathcal{T}z^\alpha\Big(-(\rho_1+\rho_2)\partial_{1 \alpha}+\rho_3 \partial_{2 \alpha}+(\rho_1+\rho_3)\partial_{\omega \alpha}\Big)\\
+\xi\left(\frac{i\rho_2}{\rho_1+\rho_2}\partial_{2 \alpha}\partial_{\go}{}^\alpha+\frac{\rho_1}{\rho_1+\rho_2}y^\alpha\partial_{\omega \alpha}\right)+(1-\xi)\Big(i(1-\rho_1)\partial_{2 \alpha}\partial_{\go}{}^\alpha-\rho_1 \partial_{1 \alpha}\partial_{\go}{}^\alpha\Big)\Big\}CC\go.
\end{multline}
Finally, one can change the integration variables to rewrite
$\widehat{W}_{2\, CC\go}^{\eta \eta}$ in the form of the integral over
a four-dimensional simplex
according to
\begin{multline}\label{4simplex}
\int_0^1 d\xi \int d^3\rho_+\, \delta\left(1-\sum_{i=1}^3 \rho_i\right)\, f(\xi,1-\xi;\rho_1,\rho_2,\rho_3)=\\
=\int d^2\xi_+\, \delta(1-\xi_1 -\xi_2)\int d^3\rho_+\, \delta\left(1-\sum_{i=1}^3 \rho_i\right)\, f(\xi_1,\xi_2;\rho_1,\rho_2,\rho_3)=\\
=\int d\zeta_1 \int d\zeta_2 \int d^2\xi_+\, \delta(1-\xi_1 -\xi_2)\int d^3\rho_+\, \delta\left(1-\sum_{i=1}^3 \rho_i\right)\,\delta(\zeta_1-\rho_3 \xi_1)\,\delta(\zeta_2-\rho_3 \xi_2) f(\xi_1,\xi_2;\rho_1,\rho_2,\rho_3)=\\
=\int d^2 \zeta_+ \int d^3 \rho_+\, \frac{1}{\rho_3^2} \delta\Big(1-\frac{\zeta_1}{\rho_3}-\frac{\zeta_2}{\rho_3}\Big)\delta\left(1-\sum_{i=1}^3 \rho_i\right)f\left(\frac{\zeta_1}{\rho_3},\frac{\zeta_2}{\rho_3};\rho_1,\rho_2,\rho_3\right)=\\
=\int d^2\zeta_+ \int d^2 \rho_+\, \frac{\delta\left(1-\rho_1-\rho_2-\zeta_1-\zeta_2\right)}{1-\rho_1-\rho_2}f\left(\frac{\zeta_1}{1-\rho_1-\rho_2},\frac{\zeta_2}{1-\rho_1-\rho_2};\rho_1,\rho_2,1-\rho_1-\rho_2\right)=\\
=\int d^4\rho_+\, \delta \left(1-\sum_{i=1}^4 \rho_i\right)\frac{1}{1-\rho_1-\rho_2}f\left(\frac{\rho_3}{1-\rho_1-\rho_2},
\frac{\rho_4}{1-\rho_1-\rho_2};\rho_1,\rho_2,1-\rho_1-\rho_2\right)\,.
\end{multline}
In these variables $\widehat{W}^{\eta\eta}_{2\, CC\go}$  acquires the form \eqref{W2CCgofinal}.

\subsection{$W_{2\, C\go C}^{\eta\eta}$}
\label{SecW2CgoC}Equation for this part of the connection is
\begin{equation}\label{W2CgoCEq}
2i \dr_z \widehat{W}_{2\, C\go C}^{\eta \eta}\approx\dr_x S_1^\eta\Big|_{C\go C}+S_1^\eta \ast W_{1\, \go C}^\eta +W_{1\, C\go}^\eta \ast S_1^\eta.
\end{equation}

Star product $S_1^\eta \ast W_{1\, \go C}^\eta$ can be computed by
\eqref{starPr}. Discarding terms in ${\Upsilon}_+^{\eta\eta}$ and
omitting the barred variables, $S_1^\eta \ast W_{1\, \go C}^\eta$
takes the form
\begin{multline}
S_1^\eta \ast W_{1\, \go C}^\eta\approx\frac{\eta^2}{2}\int_0^1 dt\int_0^1 d\tau_1 \int_0^1 d\sigma\,t(1-t)(1-\tau_1)^2 \left(\theta^\beta z_\beta\right)\left(z^\alpha \partial_{\omega \alpha}\right)\exp\Big\{i\tau_1 \circ t\, z_\alpha y^\alpha+i(1-\sigma)\partial_{\go \alpha}\partial_{2}{}^\alpha\Big\}\times\\
\times C\Big(-t(1-\tau_1)z\Big)\omega\Big(\big[\tau_1(1-t)-\sigma t(1-\tau_1)\big]z-\sigma y\Big)C\Big(\tau_1(1-t)z\Big).
\end{multline}
Since only small values of $\T$ contribute to $\widehat{\Upsilon}^{\eta\eta}$ we consider only
lower triangle of the init square in $(\tau_1,t)$  and perform the same change of variables as in
\eqref{lowTr}. Using the simplex variables \eqref{simplex1}, \eqref{simplex2} the result
 can be re-written as
\begin{multline}
S_1^\eta \ast W_{1\, \go C}^\eta\approx \frac{\eta^2}{2}\int_0^1 d\mathcal{T}\, \mathcal{T}^2\int d^3\rho_+\, \delta\left(1-\sum_{i=1}^3 \rho_i\right)\left(\theta^\beta z_\beta\right)\left(z^\alpha \partial_{\go \alpha}\right)\exp\Big\{i\mathcal{T}z_\alpha y^\alpha+i\frac{\rho_2}{\rho_1+\rho_2}\partial_{\go \alpha}\partial_2 {}^\alpha\Big\}\times\\
\times C\Big(-\mathcal{T}(1-\rho_3)z\Big)\go
\Big(-\mathcal{T}\rho_1 z+\mathcal{T}\rho_3z-\frac{\rho_1}{\rho_1+\rho_2}y\Big)C
\Big(\mathcal{T}\rho_3 z\Big)\,.
\end{multline}

Analogously, for star product $W_{1\, C\go}^\eta \ast S_1^\eta$
 \begin{multline}
W_{1\, C\go}^\eta\ast S_1^\eta\approx\frac{\eta^2}{2}\int_0^1 d\mathcal{T}\, \mathcal{T}^2\int d^3\rho_+\, \delta\left(1-\sum_{i=1}^3 \rho_i\right)\left(\theta^\beta z_\beta\right)\left(z^\alpha \partial_{\go \alpha}\right)\exp\Big\{i\mathcal{T}z_\alpha y^\alpha+i\frac{\rho_2}{\rho_1+\rho_2}\partial_{\go \alpha}\partial_2 {}^\alpha\Big\}\times\\
\times C\Big(-\mathcal{T}\rho_3 z\Big)\go\Big(\mathcal{T}\rho_1 z-\mathcal{T}\rho_3z-\frac{\rho_1}{\rho_1+\rho_2}y\Big)C\Big(\mathcal{T}(1-\rho_3) z\Big)\,.
\end{multline}

The $\dr_x S_1^\eta$ part computed using vertex \eqref{VCgoC} is
\begin{multline}
\dr_x S_1^\eta \big|_{C\go C}\approx\\
\approx-\frac{\eta^2}{2}\int_0^1 d\mathcal{T}\, \mathcal{T}^2\int d^3\rho_+\, \delta\left(1-\sum_{i=1}^3 \rho_i\right)\left(\theta^\beta z_\beta\right)\left(z^\alpha \partial_{\go \alpha}\right)\exp\Big\{i\mathcal{T}z_\alpha y^\alpha+i\rho_2\partial_{1 \alpha}\partial_{\go}{}^\alpha+i(1-\rho_2)\partial_{\go \alpha}\partial_2 {}^\alpha\Big\}\times\\
\times C\Big(-\mathcal{T}\rho_3 z\Big)\go\Big(\mathcal{T}\rho_1 z\Big)
C\Big(\mathcal{T}(1-\rho_3) z\Big)\\
-\frac{\eta^2}{2}\int_0^1 d\mathcal{T}\, \mathcal{T}^2\int d^3\rho_+\, \delta\left(1-\sum_{i=1}^3 \rho_i\right)\left(\theta^\beta z_\beta\right)\left(z^\alpha \partial_{\go \alpha}\right)\exp\Big\{i\mathcal{T}z_\alpha y^\alpha+i(1-\rho_2)\partial_{1 \alpha}\partial_{\go}{}^\alpha+i\rho_2\partial_{\go \alpha}\partial_2 {}^\alpha\Big\}\times\\
\times C\Big(-\mathcal{T}(1-\rho_3) z\Big)\go\Big(-\mathcal{T}\rho_1 z\Big)C\Big(\mathcal{T}\rho_3 z\Big)\,.
\end{multline}

It is natural to group the \rhs of (\ref{W2CgoCEq})  in the following way
\bee\nn&&
\dr_x S_1^\eta \big|_{C\go C}+S_1^\eta \ast W_{1\, \go C}^\eta+W_{1\, C\go}^\eta \ast S_1^\eta
\approx\\ \nn&&
\approx
\frac{\eta^2}{2}\int_0^1 d\mathcal{T}\, \mathcal{T}^2\int d^3\rho_+\, \delta\left(1-\sum_{i=1}^3 \rho_i\right)\left(\theta^\beta z_\beta\right)
\left(z^\alpha \partial_{\go \alpha}\right)\exp\big\{i\mathcal{T}z_\alpha y^\alpha\big\}\times
\\\nn&&\times
\Big[-\exp\big\{ i\rho_2\partial_{1 \alpha}\partial_{\go}{}^\alpha
+i(1-\rho_2)\partial_{\go \alpha}\partial_2 {}^\alpha\big\} \,\, C
\Big(-\mathcal{T}\rho_3 z\Big)\go\Big(\mathcal{T}\rho_1 z\Big)
C\Big(\mathcal{T}(1-\rho_3) z\Big)\\\nn&&
+  \exp\big\{ i\frac{\rho_2}{\rho_1+\rho_2}\partial_{\go \alpha}\partial_2 {}^\alpha\big\}\,
C\Big(-\mathcal{T}\rho_3 z\Big)\go\Big(\mathcal{T}\rho_1 z-\mathcal{T}\rho_3z-\frac{\rho_1}{\rho_1+\rho_2}y\Big)
C\Big(\mathcal{T}(1-\rho_3) z\Big)\\\nn&&
-  \exp\big\{ i(1-\rho_2)\partial_{1 \alpha}\partial_{\go}{}^\alpha
+i\rho_2\partial_{\go \alpha}\partial_2 {}^\alpha\big\}\,\, C\big(-\mathcal{T}(1-\rho_3) z\Big)
\go\Big(-\mathcal{T}\rho_1 z\Big)C\Big(\mathcal{T}\rho_3 z\Big)\\\nn&&
+  \exp\Big\{ i\frac{\rho_2}{\rho_1+\rho_2}\partial_{\go \alpha}\partial_2 {}^\alpha\big\}\,\, C\Big(-\mathcal{T}\rho_3 z\Big)\go\Big(\mathcal{T}\rho_1 z-\mathcal{T}\rho_3z-\frac{\rho_1}{\rho_1+\rho_2}y\Big)C\Big(\mathcal{T}(1-\rho_3) z\Big)
\Big]\,.\eee
Introducing new notations for brevity
\bee &&
\mathsf{Z}_1=i\mathcal{T}z_\alpha y^\alpha+\mathcal{T}z^\alpha\Big(-\rho_3 \partial_{1 \alpha}
+\rho_1 \partial_{\go \alpha}+(1-\rho_3)\partial_{2 \alpha}\Big)\\\nn&&
+\xi\left(-\mathcal{T}\rho_3 z^\alpha \partial_{\go \alpha}-\frac{\rho_1}{\rho_1+\rho_2}
y^\alpha \partial_{\go \alpha}+i\frac{\rho_2}{\rho_1+\rho_2}\partial_{1 \alpha}\partial_{\go}{}^\alpha\right)+(1-\xi)\Big(i\rho_2\partial_{1 \alpha}\partial_{\go} {}^\alpha+i(1-\rho_2)\partial_{\go \alpha}\partial_2 {}^\alpha\Big),
\\ \nn &&\mathsf{Z}_2=i\mathcal{T}z_\alpha y^\alpha+\mathcal{T}z^\alpha\Big(-(1-\rho_3)
\partial_{1 \alpha}-\rho_1 \partial_{\go \alpha}+\rho_3\partial_{2 \alpha}\Big)
\\\nn&&
+\xi\left(\mathcal{T}\rho_3 z^\alpha \partial_{\go \alpha}-\frac{\rho_1}{\rho_1+\rho_2}y^\alpha
\partial_{\go \alpha}+i\frac{\rho_2}{\rho_1+\rho_2}\partial_{\go \alpha}\partial_{2}{}^\alpha\right)+(1-\xi)
\Big(i(1-\rho_2)\partial_{1 \alpha}\partial_{\go} {}^\alpha+i\rho_2\partial_{\go \alpha}\partial_2 {}^\alpha\Big)
\,,
\eee
the \rhs of \eqref{W2CgoCEq} can be written as an integral of a total derivative
\begin{multline}\label{Z1Z2}
2i\dr_z \widehat{W}_{2\, C\go
C}^{\eta\eta}\approx\frac{\eta^2}{2}\int_0^1 d\mathcal{T}\,
\mathcal{T}^2\int d^3\rho_+\, \delta\left(1-\sum_{i=1}^3
\rho_i\right)\left(\theta^\beta z_\beta\right)\left(z^\alpha
\partial_{\go \alpha}\right)\int_0^1 d\xi \,
\frac{\partial}{\partial
\xi}\Big(e^{\mathsf{Z}_1}+e^{\mathsf{Z}_2}\Big)C\go C.
\end{multline}

Since \begin{multline}
\frac{\partial \mathsf{Z}_1}{\partial \xi}
=-\rho_3\big(\mathcal{T}z +i\partial_{1 }
+i \partial_2 {} \big){}^\alpha\partial_{\go \alpha}-\frac{\rho_1}{\rho_1+\rho_2}\Big[y^\alpha \partial_{\go \alpha}+i\rho_3 \partial_{1 \alpha}\partial_{\go} {}^\alpha+i(1-\rho_3)\partial_{\go \alpha}\partial_2 {}^\alpha\Big]=\\
=-\rho_3\big(\mathcal{T}z +i\partial_{1 }
+i \partial_2 {} \big){}^\alpha\partial_{\go \alpha}+i\frac{\rho_1}{\T(\rho_1+\rho_2)}\partial_{\go}{}^\alpha
\frac{\partial\mathsf{Z}_1}{\partial z^\alpha}\,,
\end{multline}
analogously to \eqref{SchoutenW}  by virtue of  Schouten identity  one has
\begin{multline}
\left(\theta^\beta z_\beta\right)\frac{\partial \mathsf{Z}_1}{\partial \xi}
=-\left(\theta^\beta z_\beta \right)\rho_3\big(\mathcal{T}z +i\partial_{1 }
+i \partial_2 {} \big){}^\alpha\partial_{\go \alpha}-i\frac{\rho_1
\left(z^\gamma \partial_{\go \gamma}\right)}{\mathcal{T}(\rho_1+\rho_2)}\dr_z \mathsf{Z}_1+
i\frac{\rho_1 \left(\theta^\beta \partial_{\omega \beta}\right)}
{\rho_1+\rho_2}\frac{\partial \mathsf{Z}_1}{\partial \T}\,.
\end{multline}
Therefore the $\mathsf{Z}_1$-dependent part from the \rhs of \eqref{Z1Z2}  can be rewritten in the form
\begin{multline}
\frac{\eta^2}{2}\int_0^1 d\mathcal{T}\, \mathcal{T}^2\int d^3\rho_+\, \delta\left(1-\sum_{i=1}^3 \rho_i\right)\left(\theta^\beta z_\beta\right)\left(z^\alpha \partial_{\go \alpha}\right)\int_0^1 d\xi \, \frac{\partial}{\partial \xi} e^{\mathsf{Z}_1}C\go C\approx\\
\approx-\frac{\eta^2}{2}\left(\theta^\beta z_\beta\right)\left(z^\gamma \partial_{\go \gamma}\right)
\int_0^1 d\mathcal{T}\, \mathcal{T}^2 \int_0^1 d\xi \int d^3\rho_+\, \delta\left(1-\sum_{i=1}^3 \rho_i\right)
\rho_3\big(\mathcal{T}z +i\partial_{1 }
+i \partial_2 {} \big){}^\alpha\partial_{\go \alpha}\,e^{\mathsf{Z}_1}C \go C\\
-\frac{i\eta^2}{2}\left(z^\gamma \partial_{\go \gamma}\right)^2 \int_0^1 d\mathcal{T}\, \mathcal{T} \int_0^1 d\xi\int d^3 \rho_+\, \delta\left(1-\sum_{i=1}^3 \rho_i\right)\frac{\rho_1}{\rho_1+\rho_2}\dr_z e^{\mathsf{Z}_1}C\go C\\
+\frac{i\eta^2}{2}\left(\theta^\beta \partial_{\omega
\beta}\right)\left(z^\gamma \partial_{\go \gamma}\right) \int_0^1
d\mathcal{T}\, \mathcal{T}^2 \int_0^1 d\xi\int d^3 \rho_+\,
\delta\left(1-\sum_{i=1}^3
\rho_i\right)\frac{\rho_1}{\rho_1+\rho_2}\frac{\partial}{\partial
\mathcal{T}} e^{\mathsf{Z}_1}C\go C.
\end{multline}
Integrating by parts in the last term, the resulting boundary term contributes to ${\Upsilon}_+^{\eta\eta}$.  Hence
\begin{multline}
\frac{\eta^2}{2}\int_0^1 d\mathcal{T}\, \mathcal{T}^2\int d^3\rho_+\, \delta\left(1-\sum_{i=1}^3 \rho_i\right)\left(\theta^\beta z_\beta\right)\left(z^\alpha \partial_{\go \alpha}\right)\int_0^1 d\xi \, \frac{\partial}{\partial \xi} e^{\mathsf{Z}_1}C\go C\approx\\
\approx-\frac{\eta^2}{2}\left(\theta^\beta z_\beta\right)
\left(z^\gamma \partial_{\go \gamma}\right)\int_0^1 d\mathcal{T}\, \mathcal{T}^2 \int_0^1 d\xi
\int d^3\rho_+\, \delta\left(1-\sum_{i=1}^3 \rho_i\right)\rho_3
\big(\mathcal{T}z +i\partial_{1 }
+i \partial_2 {} \big){}^\alpha\partial_{\go \alpha}\,e^{\mathsf{Z}_1}C \go C\\
+\dr_z\left\{-\frac{i\eta^2}{2}\left(z^\gamma \partial_{\go
\gamma}\right)^2 \int_0^1 d\mathcal{T}\, \mathcal{T} \int_0^1
d\xi\int d^3 \rho_+\, \delta\left(1-\sum_{i=1}^3
\rho_i\right)\frac{\rho_1}{\rho_1+\rho_2} e^{\mathsf{Z}_1}C\go C
\right\}.
\end{multline}
Analogously, since
\be\nn
\frac{\partial \mathsf{Z}_2}{\partial \xi}
 =\rho_3\big(\mathcal{T}z +i\partial_{1 }
+i \partial_2 {} \big){}^\alpha\partial_{\go \alpha}
+i\frac{\rho_1}{\T(\rho_1+\rho_2)}\partial_{\go} {}^\alpha \frac{\partial \mathsf{Z}_2}{\partial z^\alpha}\,,
\ee
\begin{multline}
\left(\theta^\beta z_\beta\right)\frac{\partial \mathsf{Z}_2}{\partial \xi}=
\left(\theta^\beta z_\beta\right)\rho_3
\big(\mathcal{T}z +i\partial_{1 }
+i \partial_2 {} \big){}^\alpha\partial_{\go \alpha}-i\frac{\rho_1
\left(z^\gamma \partial_{\go\gamma}\right)}{\T(\rho_1+\rho_2)}\dr_z \mathsf{Z}_2+
 i\frac{\rho_1\left(\theta^\beta \partial_{\omega \beta}\right)}{\rho_1
+\rho_2}\frac{\partial \mathsf{Z}_2}{\partial \T}
\end{multline} and,
therefore, the $\mathsf{Z}_2$-dependent part of the r.h.s. of \eqref{Z1Z2} yields
\begin{multline}
\frac{\eta^2}{2}\int_0^1 d\mathcal{T}\, \mathcal{T}^2\int d^3\rho_+\, \delta\left(1-\sum_{i=1}^3 \rho_i\right)\left(\theta^\beta z_\beta\right)\left(z^\alpha \partial_{\go \alpha}\right)\int_0^1 d\xi \, \frac{\partial}{\partial \xi} e^{\mathsf{Z}_2}C\go C\approx\\
\approx\frac{\eta^2}{2}\left(\theta^\beta z_\beta\right)\left(z^\gamma \partial_{\go \gamma}\right)
\int_0^1 d\mathcal{T}\, \mathcal{T}^2 \int_0^1 d\xi \int d^3\rho_+\, \delta\left(1-\sum_{i=1}^3 \rho_i\right)
\rho_3\big(\mathcal{T}z +i\partial_{1 }
+i \partial_2 {} \big){}^\alpha\partial_{\go \alpha}\,e^{\mathsf{Z}_2}C \go C\\
-\frac{i\eta^2}{2}\left(z^\gamma \partial_{\go \gamma}\right)^2 \int_0^1 d\mathcal{T}\, \mathcal{T}
\int_0^1 d\xi\int d^3 \rho_+\, \delta\left(1-\sum_{i=1}^3 \rho_i\right)\frac{\rho_1}{\rho_1+\rho_2}
\dr_z e^{\mathsf{Z}_2}C\go C\\
+\frac{i\eta^2}{2}\left(\theta^\beta \partial_{\omega
\beta}\right)\left(z^\gamma \partial_{\go \gamma}\right) \int_0^1
d\mathcal{T}\, \mathcal{T}^2 \int_0^1 d\xi\int d^3 \rho_+\,
\delta\left(1-\sum_{i=1}^3
\rho_i\right)\frac{\rho_1}{\rho_1+\rho_2}\frac{\partial}{\partial
\mathcal{T}} e^{\mathsf{Z}_2}C\go C\,.
\end{multline}
Modulo terms from $\mathcal{H}^+$ this equals to
 \begin{multline}
  \frac{\eta^2}{2}\left(\theta^\beta z_\beta\right)\left(z^\gamma \partial_{\go \gamma}\right)
  \int_0^1 d\mathcal{T}\, \mathcal{T}^2 \int_0^1 d\xi \int d^3\rho_+\, \delta\left(1-\sum_{i=1}^3 \rho_i\right)
  \rho_3\big(\mathcal{T}z +i\partial_{1 }
+i \partial_2 {} \big){}^\alpha\partial_{\go \alpha}\,e^{\mathsf{Z}_2}C \go C\\
+\dr_z\left\{-\frac{i\eta^2}{2}\left(z^\gamma \partial_{\go
\gamma}\right)^2 \int_0^1 d\mathcal{T}\, \mathcal{T} \int_0^1
d\xi\int d^3 \rho_+\, \delta\left(1-\sum_{i=1}^3
\rho_i\right)\frac{\rho_1}{\rho_1+\rho_2} e^{\mathsf{Z}_2}C\go C
\right\}.
\end{multline}
As a result, the \rhs of \eqref{Z1Z2} acquires the form

 \begin{multline}
2i\dr_z \widehat{W}_{2\, C\go C}^{\eta \eta}\approx\dr_z\left\{-\frac{i\eta^2}{2}\left(z^\gamma \partial_{\go \gamma}\right)^2 \int_0^1 d\mathcal{T}\, \mathcal{T} \int_0^1 d\xi\int d^3 \rho_+\, \delta\left(1-\sum_{i=1}^3 \rho_i\right)\frac{\rho_1}{\rho_1+\rho_2}\Big[e^{\mathsf{Z}_1}+ e^{\mathsf{Z}_2}\Big]C\go C \right\}\\
+\frac{\eta^2}{2}\left(\theta^\beta z_\beta\right)\left(z^\gamma \partial_{\go \gamma}\right)
\int_0^1 d\mathcal{T}\, \mathcal{T}^2 \int_0^1 d\xi \int d^3\rho_+\,
\delta\left(1-\sum_{i=1}^3 \rho_i\right)\rho_3
\big(\mathcal{T}z +i\partial_{1 }
+i \partial_2 {} \big){}^\alpha\partial_{\go \alpha}\left[e^{\mathsf{Z}_2}-e^{\mathsf{Z}_1}\right]C \go C .
\end{multline}
To see that the last   term  vanishes it is convenient to change
integration variables to those of the four-dimensional simplex
\eqref{4simplex}. In these four-dimensional simplicial variables
 \begin{multline}\label{Z1=}
\mathsf{Z}_1=i\mathcal{T}z_\alpha y^\alpha+\mathcal{T}z^\alpha\Big(-(\rho_3+\rho_4)\partial_{1 \alpha}+(\rho_1-\rho_3)\partial_{\go \alpha}+(\rho_1+\rho_2)\partial_{2 \alpha}\Big)
-\frac{\rho_3\rho_1}{(\rho_1+\rho_2)(\rho_3+\rho_4)}y^\alpha \partial_{\go\alpha}\\+i\frac{\rho_3\rho_2}{(\rho_1+\rho_2)(\rho_3+\rho_4)}\partial_{1 \alpha}\partial_{\go} {}^\alpha+i\frac{\rho_4 \rho_2}{\rho_3+\rho_4}\partial_{1 \alpha}\partial_{\go} {}^\alpha+i\frac{\rho_4(1-\rho_2)}{\rho_3+\rho_4}\partial_{\go \alpha}\partial_2 {}^\alpha\,,
\end{multline}

\begin{multline}\label{Z2=}
\mathsf{Z}_2=i\mathcal{T}z_\alpha y^\alpha+\mathcal{T}z^\alpha\Big(-(\rho_1+\rho_2)\partial_{1 \alpha}+(\rho_3-\rho_1)\partial_{\go \alpha}+(\rho_3+\rho_4)\partial_{2 \alpha}\Big)
-\frac{\rho_3\rho_1}{(\rho_1+\rho_2)(\rho_3+\rho_4)}y^\alpha \partial_{\go\alpha}\\
+i\frac{\rho_3\rho_2}{(\rho_1+\rho_2)(\rho_3+\rho_4)}\partial_{\go \alpha}\partial_{2} {}^\alpha+i\frac{\rho_4 \rho_2}{\rho_3+\rho_4}\partial_{\go \alpha}\partial_{2} {}^\alpha+i\frac{\rho_4(1-\rho_2)}{\rho_3+\rho_4}\partial_{1 \alpha}\partial_\go {}^\alpha\,.
\end{multline}
Shuffling the $\rho$-variables in $\mathsf{Z}_2$
\begin{equation}
\rho_1\longrightarrow\rho_3,\;\; \rho_3\longrightarrow \rho_1,\;\; \rho_2\longrightarrow\rho_4,\;\;
\rho_4\longrightarrow\rho_2\q
\end{equation}
taking into account that
\bee\nn
&&\gd(1-\rho_1-\rho_2-\rho_3-\rho_4)\left(\frac{\rho_1\rho_4}{(\rho_1+\rho_2)(\rho_3+\rho_4)}
+\frac{\rho_4\rho_2}{\rho_1+\rho_2}\right) =
\gd(1-\rho_1-\rho_2-\rho_3-\rho_4) \frac{\rho_4(1-\rho_2)}{\rho_3+\rho_4},
\\\nn&& \gd(1-\rho_1-\rho_2-\rho_3-\rho_4)\left(\frac{\rho_3\rho_2}{(\rho_1+\rho_2)(\rho_3+\rho_4)}
+\frac{\rho_4\rho_2}{\rho_3+\rho_4}\right)
   =\gd(1-\rho_1-\rho_2-\rho_3-\rho_4)\frac{\rho_2(1-\rho_4)}{\rho_1+\rho_2}\q
\eee
one can  see that
\begin{multline}
\mathsf{Z}_1=\mathsf{Z}_2=i\mathcal{T}z_\alpha y^\alpha+\mathcal{T}z^\alpha\Big(-(\rho_3+\rho_4)\partial_{1 \alpha}+(\rho_1-\rho_3)\partial_{\go \alpha}+(\rho_1+\rho_2)\partial_{2 \alpha}\Big)
\\-\frac{\rho_3\rho_1}{(\rho_1+\rho_2)(\rho_3+\rho_4)}y^\alpha \partial_{\go\alpha}
+i\frac{\rho_4(1-\rho_2)}{\rho_3+\rho_4}\partial_{\go \alpha}\partial_{2} {}^\alpha+i\frac{\rho_2(1-\rho_4)}{\rho_1+\rho_2}\partial_{1 \alpha}\partial_\go {}^\alpha.
\end{multline}

The part of $W_{2\, C\go C}^{\eta\eta}$ that contributes to $\widehat{\Upsilon}^{\eta\eta}$
thus has the form \eqref{W2CgoCfinal}.

\subsection{$W_{2\, \go CC}^{\eta\eta}$}
Equation for the part of $W_{2\, \go CC}^{\eta\eta}$ that contributes to $\widehat{\Upsilon}^{\eta\eta}$ is
\begin{equation}
2i\dr_z \widehat{W}_{2\, \go CC}^{\eta\eta}\approx\dr_x S_1^\eta \big|_{\go CC}+W_{1\, \go C}^\eta\ast S_1^\eta.
\end{equation}
Since computation is analogous to that for $\widehat{W}_{2\ CC\go}^{\eta\eta}$ we present only the final result
\begin{multline}
\widehat{W}_{2\, \omega CC}^{\eta\eta}=\frac{\eta^2}{4}\int_0^1 d\mathcal{T} \, \mathcal{T}\int_0^1 d\xi\int d^3\rho_+\, \delta\left(1-\sum_{i=1}^3 \rho_i\right)\frac{\rho_1 (z^\alpha \partial_{\omega\alpha})^2}{\rho_1+\rho_2}\times\\
\times\exp\bigg\{i\mathcal{T}z_\alpha y^\alpha+\mathcal{T}z^\alpha\Big((\rho_1+\rho_3) \partial_{\omega\alpha}-\rho_3 \partial_{1\alpha}+(\rho_1+\rho_2) \partial_{2\alpha}\Big)\\
+\xi\Big(i(1-\rho_1)\partial_{\omega\alpha}\partial_1 {}^\alpha-i\rho_1\partial_{\omega\alpha}\partial_2 {}^\alpha\Big)+(1-\xi)\left(i\frac{\rho_2}{\rho_1+\rho_2}\partial_{\omega\alpha}\partial_1 {}^\alpha+\frac{\rho_1}{\rho_1+\rho_2}y^\alpha \partial_{\omega \alpha}\right)\bigg\}\omega CC.
\end{multline}
Equivalently, as an integral over a four-dimensional simplex it is given in \eqref{W2goCCfinal}.

\addcontentsline{toc}{section}{\,\,\,\,\,\,\,References}
\section*{}

\end{document}